\documentclass[11pt,draftcls,onecolumn,letter]{IEEEtran}

\usepackage{amsmath} 
\usepackage{amssymb}  
\usepackage{latexsym}
\usepackage{graphicx}
\usepackage{subfigure}

\newtheorem{definition}{Definition}

\newtheorem{theorem}{Theorem}

\newtheorem{fact}[definition]{Fact}
\newtheorem{example}[definition]{Example}
\newtheorem{problem}{Problem}

\newcommand{\Proof}{\noindent\mbox{\bf Proof} \hspace{0.3cm}}
\newcommand{\EndProof}{\hfill \mbox{$\Box$}}

\newcommand{\Real}{\mathop{\rm Re}\nolimits}

\newcommand{\esssup}{\mathop{\rm esssup}}

\newcommand{\R}{{\mathbb R}}

\newcommand{\C}{{\mathbb{C}}}
\newcommand{\D}{{\mathbb{D}}}

\newcommand{\Hinf}{H^\infty}

\newcommand{\Hinfnorm}[1]{\left\|#1\right\|_\infty}
\newcommand{\Twonorm}[1]{\left\|#1\right\|_2}

\newcommand{\innprod}[2]{\mbox{$\left<#1, #2\right>$}}

\newcommand{\diag}{\mathop{\rm diag}}

\newcommand{\dmat}[4]{\left[ 
    \begin{array}{cc}
      #1&#2\\
      #3&#4
    \end{array}\right]}

\newcommand{\ssr}[4]{\left[ 
    \begin{array}{c|c}
      #1&#2\\ \hline 
      #3&#4
    \end{array}
  \right]}

\newcommand{\hold}[1]{{\cal H}_{#1}}
\newcommand{\ghold}[1]{\widetilde{\cal H}_{#1}}

\newcommand{\samp}[1]{{\cal S}_{#1}}

\newcommand{\us}[1]{\uparrow\!#1}
\newcommand{\ds}[1]{\downarrow\!#1}

\newcommand{\dlift}[1]{{\bf L}_{#1}}
\newcommand{\idlift}[1]{{\bf L}_{#1}^{-1}}
\newcommand{\Gd}{\dmat{z^{-m}F_N(z)}{\!-P_N(z)H}{SF_N(z)}{0}}
\newcommand{\Fa}{F_{\text{a}}}
\newcommand{\fa}{f}

\begin{document}

\title{Signal Reconstruction via $H^{\infty}$ Sampled-Data Control 
Theory---Beyond the Shannon Paradigm}

\author{
\thanks{
Copyright (c) 2011 IEEE. Personal use of this material is permitted. 
However, permission to use this material for any other purposes must be obtained 
from the IEEE by sending a request to pubs-permissions@ieee.org.
}
Yutaka~Yamamoto\thanks{Y.~Yamamoto is with 
Department of Applied Analysis and Complex Dynamical Systems, 
Graduate School of Informatics, Kyoto University, Kyoto, 606-8501, JAPAN 
    (e-mail: {\tt yy@i.kyoto-u.ac.jp}).
    Mailing address: Kyoto University, Yoshida Honmachi, Sakyo-ku, Kyoto 606-8501, Japan.  TEL: +81-75-753-5901, FAX: +81-75-753-5517.},~\IEEEmembership{Fellow,~IEEE}, 
Masaaki~Nagahara\thanks{M.~Nagahara is with 
Department of Applied Analysis and Complex Dynamical Systems,
Graduate School of Informatics, Kyoto University, Kyoto, 606-8501, JAPAN 
    (e-mail: {\tt nagahara@i.kyoto-u.ac.jp}).
    Mailing address: Kyoto University, Yoshida Honmachi, Sakyo-ku, Kyoto 606-8501, Japan.  TEL: +81-75-753-5903.},~\IEEEmembership{Member,~IEEE},
and\\Pramod P. Khargonekar\thanks{P. P. Khargonekar is with Department of Electrical and Computer Engineering, University of Florida, 
Gainesville, FL 32611-620, U. S. A.
(e-mail: {\tt ppk@ufl.edu})
Mailing address: 427 New Engineering Building, 
University of Florida, Gainesville, FL 32611-6130, U. S. A. 
TEL: +1-352-392-0918.},~\IEEEmembership{Fellow,~IEEE}}%

\maketitle

\begin{abstract}
  This paper presents a new method for signal reconstruction by leveraging 
sampled-data control theory. We formulate the signal reconstruction problem 
in terms of an  analog performance optimization problem using a stable discrete-time filter. 
The proposed $H^{\infty}$ performance criterion naturally takes
intersample behavior into account, reflecting the energy
distributions of the signal.
We present methods for computing optimal solutions which are guaranteed to be stable and causal.
Detailed comparisons to alternative methods are provided.  
We discuss some applications in sound and image reconstruction.
\end{abstract}

%

\section{Introduction}\label{Section:introduction}

\PARstart{S}{ignal} reconstruction from digital data is at the foundations of 
 digital signal processing.  For given digital data, stored or 
transmitted, we need to recover the original 
signal that generated the data.  
This procedure is needed and used everywhere: 
image and sound reconstruction/restoration, moving images, 
mobile telephones, etc.  While discussion of transmission/recovery for 
digital data only is quite routine, we should note that our ultimate objective
is to recover or reconstruct the original {\em analog\/} data
from which such digital data are generated.

This signal reconstruction problem dates back to the celebrated 
paper of Shannon \cite{Shannon1}.  
Using the sampling theorem \cite{Zayed}, he showed that  
we can recover the original analog information from 
sampled data, provided that the original 
analog signal is band-limited below the Nyquist frequency.  
Based on this result, he established a fundamental paradigm 
for communication and digital processing.  
We will hereafter refer to this scheme {\em the Shannon paradigm}. 

This paradigm however leads to various unrealistic constraints.  
The assumption of perfect band-limitedness, necessary for 
perfect signal reconstruction, is hardly satisfied in 
reality.  In many applications, the sampling rate is
not high enough to allow for this assumption to hold even
approximately.  To remedy this drawback, one often introduces 
an anti-aliasing filter to sharply cut  high frequency components, 
but this in turn leads to yet another type of distortion 
due to the Gibbs phenomenon 
(see Example \ref{Ex:Johnston-yy} in Section \ref{Sec:FSFH} 
and also Section \ref{Sec:Appl} below).  
Secondly, the sinc function, which is the impulse response
of the ideal filter, is not causal and does not decay very fast.  
This slow decay rate makes it very difficult to implement and various approximations (mostly with
respect to the $H^{2}$-norm criterion) become necessary.
This procedure further complicates the total design procedure,
making it less transparent. 

In view of such drawbacks, there has been revived interest in 
the extension of the sampling theorem in various forms since the 1990's.  
There is by now a stream of papers that aim to 
study signal reconstruction under the assumption of 
non-ideal signal acquisition devices; an excellent 
survey is given in \cite{Unser00}.  
In this research framework, 
 the incoming signal is 
acquired through a non-ideal 
analog filter (acquisition device) and sampled, 
and then the reconstruction process attempts to
recover the original signal.
The idea is to place the problem into the framework 
of the (orthogonal or oblique) 
projection theorem in a Hilbert space (usually $L^{2}$), 
and then project the signal space to the subspace generated by 
the shifted reconstruction functions.  
It is often required that the 
process give a {\em consistent\/} result, i.e., 
if we subject the reconstructed signal to the
whole process again, it should yield the same 
sampled values from which it was reconstructed \cite{Unser94}.

The objective of the present paper is to go beyond the 
Shannon paradigm, and present an entirely new scheme of 
signal reconstruction, different from the ones described above.  
For our scheme, we draw upon modern 
sampled-data control theory, developed in the control community
since the 1990's. The fundamental accomplishment of modern sampled-data control
theory is that it can give us a discrete-time 
controller (or filter) 
that optimizes the closed-loop performance with intersample 
behavior taken into account.  In other words, it can optimize 
an {\em analog (continuous-time) performance}.  Such a performance 
is measured according to the $H^{\infty}$ or $H^{2}$ performance 
criteria. This setting gives us an optimal platform 
to reconstruct the original analog signals from sampled-data
under the scenario that the original signal is not
band-limited below the Nyquist frequency.

Chen and Francis \cite{ChenFrancissignal} 
made a first attempt to apply sampled-data control theory
to signal processing (however in a 
discrete-time domain); see also \cite{Hassibi2006}.  
Starting in 1995,
the present authors and our colleagues have pursued the signal reconstruction problem
 in the sampled-data context to obtain an optimal 
analog performance via digital filtering: See 
\cite{PPKYYCDC95,YYMTNS2000,YYAust05} for general design frameworks, 
\cite{IshiiYYFrancis99,NagYYCDC00} for sample-rate conversion,  
\cite{YYFujPPK97} for multirate filterbank design, 
\cite{AshidaNagYY03,AshidaKakNagYY04} for audio signal compression, 
\cite{KakNagKobYY05} for image restoration, 
\cite{NagYY05} for fractional delay filters,  
\cite{KashimaYYNag04} for wavelet expansion, and
\cite{YYAutoma99,YYCDC02} for convergence analysis

The basic approach is as follows:
We start with a signal generation model 
that consists of an analog filter with $L^{2}$ inputs. 
Then we formulate the signal reconstruction problem as a 
sampled-data $H^{\infty}$ control design problem.  The controller 
to be designed is the digital filter that is desired to 
reconstruct the original analog signal.  The advantage here 
is that we can formulate an {\em overall error system}, and be able
to have control over all frequencies including {\em both gain 
and phase errors}, not merely the gain characteristics often 
observed in many filter designs.  Introducing an upsampler, this
framework also enables us to optimally interpolate the 
{\em intersample high frequency components\/} based on the
model of the signal generator.  We will formulate and discuss 
this in more detail in Section \ref{Sec:SignalReconstruction}.  

The same philosophy of emphasizing the importance 
of analog performance was proposed and pursued 
recently by Unser and co-workers \cite{UnserBlu05,Unser05}. 
The crucial difference is however that 
they rely on $L^{2}/H^{2}$ type optimization and oblique projections, 
which are very different from our method here.  In particular, 
it can raise some stability questions.  
The recent work of Meinsma and Mirkin 
\cite{MeinsmaMirkinSP10a,MeinsmaMirkinSP10b} takes an approach that is close to ours. 
They give solutions for non-causal 
problems and allow freedom in the choice of sample or hold devices.
A detailed comparison of our work and these related works is 
provided in Section \ref{Sec:Comparison}.
Some other approaches (not very closely related to our work)
to extending the traditional sampling theory include:
reconstruction by quasi-interpolation \cite{CondatBluUnser2005},
and minimization of the worst-case regret \cite{EldarDvorkind2006}.

The present paper is organized as follows:  
After preparing some basic notions in function spaces in 
Section \ref{Sec:preliminaries}, we first review the 
fundamentals in signal reconstruction using the sampling 
theorem, and discuss its various drawbacks in 
Section \ref{Sec:samplingthm}.  We will then give a 
fundamental setup and formulation of our sampled-data 
filter design framework in Section \ref{Sec:SignalReconstruction}, 
and Section \ref{Sec:FSFH} gives a solution method via
fast-sample/fast-hold approximation. 
In Section \ref{Sec:Comparison}, 
we discuss some related work and make comparison with 
the methods and results proposed by Unser and co-workers
\cite{Unser05,UnserBlu05} and also those of 
Meinsma and Mirkin \cite{MeinsmaMirkinSP10a,MeinsmaMirkinSP10b}. 
Finally, we give some examples in signal reconstruction 
for sounds and images in Section \ref{Sec:Appl}.

\section{Preliminaries}
\label{Sec:preliminaries}

Let us introduce some basic function spaces and performance
measures.  Let $L^{2}(a, b)$ (or $L^{2}[a, b)$, $L^{2}[a, b]$, etc.) 
be the space of Lebesgue square integrable functions on the 
interval $(a, b)$, $a < b$.  
For a function $f$ valued in $\R^{n}$ or $\C^{n}$, its $L^{2}$-norm
is denoted by 
\begin{equation} \label{eqn:twonorm}
  \Twonorm{f} = \left\{\int_{a}^{b} |f(t)|^{2}dt \right\}^{1/2}, 
\end{equation}
where $|\cdot|$ denotes the Euclidean norm in $\C^{n}$.  
Let $H^{2}$ denote the space of $\C^{n}$-valued functions $f$
that are analytic on the open right half plane 
$\C_{+} := \{s : \Real s > 0 \}$ and satisfy 
\[
      \sup_{x > 0}\int_{-\infty}^{\infty} 
                    |f(x+jy)|^{2}dy < \infty.  
\]
The $H^{2}$-norm of a function $f\in H^{2}$ is defined by 
\begin{equation} \label{eqn:H2norm}
 \Twonorm{f} := \sup_{x > 0} \left\{\frac{1}{2\pi}
                   \int_{-\infty}^{\infty} 
                    |f(x+jy)|^{2}dy
               \right\}^{1/2}
\end{equation}
It is well known that Laplace transform gives an isometry
between $L^{2}[0, \infty)$ and $H^{2}$.  

The space $\Hinf$ denotes the Hardy space of functions 
analytic on $\C_{+}$ and bounded there.  It is a Banach space
with norm
\begin{equation} \label{eqn:Hinfnorm}
  \Hinfnorm{f} := \sup_{s \in \C_{+}} |f(s)|.  
\end{equation}
An element $f$ of $\Hinf$ admits nontangential limit to the 
imaginary axis almost everywhere, 
which we denote by $f(j\omega)$, $\omega \in \R$.  
Then the $H^{\infty}$-norm of $f \in \Hinf$ is equal to
\begin{equation} \label{eqn:Bodemaximum}
  \Hinfnorm{f} = \esssup_{-\infty < \omega < \infty} |f(j\omega)|. 
\end{equation}

Now let $G$ be the transfer function 
of a finite-dimensional, asymptotically stable 
linear continuous-time system.  Then $G$ 
belongs to $H^{\infty}$, and its ``size'' is measured by the 
$H^{\infty}$-norm, i.e., the supremum (or maximum) of the 
Bode magnitude plot as in (\ref{eqn:Bodemaximum}).  

The steady-state response of $G$ against 
a sinusoid $e^{j\omega t}$ is given by 
$G(j\omega)e^{j\omega t}$, and its magnitude is bounded as 
\[
  |G(j\omega)e^{j\omega t}| \leq \sup_{-\infty < \omega < \infty} |G(j\omega)|
                \cdot |e^{j\omega t}| = \Hinfnorm{G}.
\]
In general, for $u \in H^{2}$, it is known that 
\begin{equation}\label{eqn:Hinfnorm-ieq}
  \Twonorm{Gu} \leq \Hinfnorm{G}\cdot\Twonorm{u}, 
\end{equation}
and this bound is tight.  
Hence the $H^{\infty}$ norm gives the $L^{2}$ energy induced-gain, 
and minimizing it yields a system that works uniformly 
well for the whole frequency range%
\footnote{
However, it is to be noted that it is not possible to uniformly
attenuate $|G(j\omega)|$.  If we attenuate $G(j\omega)$ for a 
certain frequency range, it will yield an amplification 
at another range.  Due to this effect, one usually introduce 
a frequency weighting $W(s)$, and minimize $|W(j\omega)G(j\omega)|$.}.

For this reason, it is recognized that the $\Hinf$-norm criterion 
is often superior to the $H^{2}$-norm criterion, 
where the $H^{2}$-norm for a stable matrix transfer function is defined as
\[
 \Twonorm{G} := \left(\frac{1}{2\pi}\int_{-\infty}^\infty 
       \mbox{trace }\{G^{*}(j\omega)G(j\omega)\} dt\right)^{1/2}. 
\]
The $H^\infty$ norm has been used successfully in the control literature 
\cite{DFT,DGKF,Francisbook}.

The $H^\infty$ norm criterion is naturally extended to sampled-data systems.  
The problem here is that such systems have two time sets: continuous and
discrete.  Hence the overall system is not time-invariant in the 
classical sense.  This difficulty can be remedied by the 
now-standard technique called {\em lifting,} which converts a
linear time-invariant continuous-time system to an 
infinite-dimensional discrete-time system. 
It is then possible to naturally extend the notion of the 
$H^{\infty}$-norm to sampled-data systems.  To be more precise,
let ${\mathcal G}$ denote the input/output operator of such a 
system.  Then its $H^{\infty}$-norm is defined to be the 
induced norm against all $L^{2}$ inputs: 
\begin{equation} \label{eqn:SDinducednorm}
 \Hinfnorm{{\mathcal G}} := 
            \sup_{u\in L^2,u\neq 0}\frac{\|{\mathcal G}u\|_2}{\|u\|_2}.
\end{equation} 
Via {\em lifting}, this norm is shown to be equivalent to
the maximum gain of the frequency response operator of ${\mathcal G}$
as in (\ref{eqn:Hinfnorm}).  
For details, see Appendix \ref{Sec:appendLifting}.

\section{Signal reconstruction and sampling theorem}
\label{Sec:samplingthm}

Consider the block diagram depicted in Fig.~\ref{fig:sigprocess}:

\noindent
\begin{figure}[h]
\begin{picture}(9.1,1)(-1,0)
\unitlength 0.8cm
\put(0,0.5){\vector(1,0){1}}
\put(0.5,0.75){\makebox(0,0)[b]{$ w_{c} $}}
\put(1,0){\framebox(1,1){$ F $}}
\put(2,0.5){\line(1,0){0.5}}
\put(2.35,0.75){\makebox(0,0)[b]{$ y_{c} $}}
\put(2.5,0.5){\line(2,1){0.6}}
\put(3.5,0.75){\makebox(0,0)[b]{$ y_{d} $}}
\put(3.1,0.5){\vector(1,0){1}}
\put(4.1,0){\framebox(1.5,1){$ K(z) $}}
\put(5.6,0.5){\vector(1,0){1}}
\put(6.1,0.75){\makebox(0,0)[b]{$ y_{K,d} $}}
\put(6.6,0){\framebox(1.5,1){$ \Phi $}}
\put(8.1,0.5){\vector(1,0){1}}
\put(8.6,0.75){\makebox(0,0)[b]{$ y $}}
\end{picture}
\caption{Signal Reconstruction System}
\label{fig:sigprocess}
\end{figure}
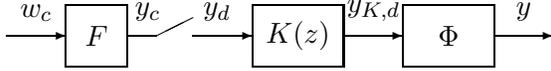

In this diagram, the signal $w_{c} \in L^{2}$ denotes the external analog 
signal to be reconstructed.  It is filtered by an analog filter 
(acquisition device) $F$, and then sampled by the
sampler with sampling period $h$.  If $f(t)$ denotes 
the impulse response of the analog filter $F$, then the 
discrete-time signal $y_{d}[k]$ is easily seen to be given by 
\begin{equation} \label{eqn:sampled}
  y_{d}[k] = (f*w_{c})[k] = \innprod{\check{f}(\cdot - kh)}{w_{c}}
\end{equation}
where $\check{f}(t) := f(-t)$ is the mirror image 
(with respect to time) of $f$,
and $\langle f,g \rangle$ denotes the inner product in $L^2$.
The obtained signal $y_{d}$ is then processed by a discrete-time 
filter $K$ and then the filtered discrete-time signal 
$y_{K,d}$ is converted back to an analog signal $y$ via a reconstruction 
device $\Phi$.  Denoting by $\phi$ the impulse response of 
$\Phi$, the reconstructed $y$ is given by 
\begin{equation} \label{eqn:reconstruction}
  y(t) = \sum_{k=-\infty} ^{\infty} y_{K,d}[k] \phi(t - kh).
\end{equation}

In the Shannon paradigm, the analog filter $F$ is taken to be the 
ideal filter, and $\phi$ above is the {\em sinc\/} function 
\cite{Zayed,Unser00}.  
As mentioned in the Introduction, this has several limitations.  
To take care of this, one often employs an 
approximation of the ideal filter with respect to $H^{2}$ norm
\cite{Fliege}, and this unfortunately 
yields a sharp ringing effect in the frequency domain.  

Unser and co-workers published series of 
papers of generalized sampling theorems where the
acquisition device $F$ is not the ideal filter 
\cite{Unser94,UnserZerubia97,Unser00}.  
First define the subspace
\begin{equation} \label{eqn:generationspace}
  V_{f} := 
     \{ \sum_{k=-\infty} ^{\infty} \alpha[k] f(t - kh) : 
                      \{ \alpha[k]\} \in \ell^{2} \} 
\end{equation}
generated by the translates of the impulse response of 
the acquisition filter, and the reconstruction space 
\begin{equation} \label{eqn:reconstructionsp}
  V_{\phi} := \{ \sum_{k=-\infty} ^{\infty} \beta[k] \phi(t - kh) : 
                            \{ \beta[k]\} \in \ell^{2}   \}
\end{equation}
generated by the translates of the reconstruction function $\phi$.  
From the consistency requirement \cite{Unser94}, 
a key step in their procedure is 
the oblique projection of $L^{2}$ onto $V_{\phi}$ perpendicular 
to $V_{f}$.  
A precise comparison of this approach with our work is given in 
Section \ref{Sec:Comparison}.

\section{$H^{\infty}$ Signal Reconstruction Problem} 
\label{Sec:SignalReconstruction}

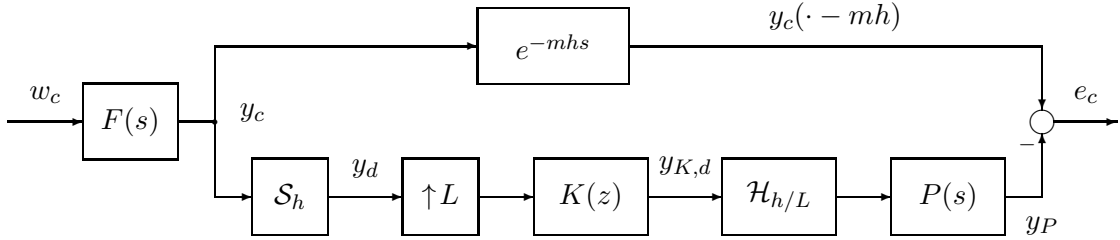
\begin{figure}[tb]
\centering
\begin{picture}(23,5)(0,0)
\typeout{}\typeout{Hello!}\typeout{}			
\unitlength 1.0cm
\put(0,1.5){\vector(1,0){1}}
\put(0.5,1.75){\makebox(0,0)[b]{$ w_{c} $}}
\put(1,1){\framebox(1.25,1){$ F(s) $}}
\put(2.25,1.5){\line(1,0){0.5}}
\put(2.75,1.5){\circle*{0.08}}
\put(3.25,1.50){\makebox(0,0)[b]{$ y_{c} $}}
\put(2.75,1.5){\line(0,1){1}}
\put(2.75,2.5){\vector(1,0){3.5}}
\put(6.25,2){\framebox(2,1){$ e^{-mhs} $}}
\put(8.25,2.5){\line(1,0){5.5}}
\put(11,2.70){\makebox(0,0)[b]{$ y_{c}(\cdot - mh) $}}
\put(13.75,2.5){\vector(0,-1){0.85}}
\put(13.75,1.5){\circle{0.3}}
\put(13.9,1.5){\vector(1,0){0.85}}
\put(14.33,1.75){\makebox(0,0)[b]{$ e_{c} $}}
\put(2.75,1.5){\line(0,-1){1}}
\put(2.75,0.5){\vector(1,0){0.5}}
\put(3.25,0){\framebox(1,1){$ {\cal S}_{h} $}}
\put(4.25,0.5){\vector(1,0){1}}
\put(4.75,0.75){\makebox(0,0)[b]{$ y_{d} $}}
\put(5.25,0){\framebox(1,1){$ \us L $}}
\put(6.25,0.5){\vector(1,0){0.75}}
\put(7,0){\framebox(1.5,1){$ K(z) $}}
\put(8.5,0.5){\vector(1,0){1}}
\put(9,0.75){\makebox(0,0)[b]{$ y_{K,d} $}}
\put(9.5,0){\framebox(1.5,1){$ \hold{h/L} $}}
\put(11,0.5){\vector(1,0){0.75}}
\put(11.75,0){\framebox(1.5,1){$ P(s) $}}
\put(13.25,0.5){\line(1,0){0.5}}
\put(13.75,0){\makebox(0,0)[b]{$ y_{P} $}}
\put(13.75,0.5){\vector(0,1){0.85}}
\put(13.55,1.2){\makebox(0,0){$ \scriptstyle - $}}
\end{picture}
\caption{Error system of a sampled-data design filter}
\label{Fig:err_interp}
\end{figure}

We are now ready to precisely state our signal reconstruction
problem.  The basic features are the following:
\begin{itemize}
\item We allow a finite step preview for reconstruction.
\item The acquisition device, sampling and hold elements are fixed.  
\end{itemize}
Consider the block diagram Fig.~\ref{Fig:err_interp}. 
The external continuous-time signal $w_{c} \in L^{2}$ 
is first
filtered, or  band-limited (mildly but not perfectly)
by going through the analog low-pass filter 
$F(s)$, which is linear and time-invariant, and finite-dimensional.  
This $F(s)$ is a rational function 
of $s$ which is strictly proper (i.e., the degree 
of the numerator polynomial is less than that of 
the denominator).  
As is well known, it is represented 
by a linear, time-invariant system
\begin{eqnarray*}
 \frac{dx}{dt}(t) & = & Ax(t) + Bu(t) \\
  y(t) & = & Cx(t)
\end{eqnarray*}
where $A, B, C$ are constant matrices of appropriate
sizes, and $F(s) = C(sI - A)^{-1}B$.  Hence $F$ is not 
an ideal filter unlike the case of the Shannon paradigm, 
and is physically realizable through the above state space model.
The signal $w_{c}$ is the external signal that drives 
$F$ and produces the actual signal $y_{c}$ to be processed.  
That is, we assume that the original analog signals to be sampled
are in the following subspace of $L^2$:
\[
 FL^2 := \left\{y_c\in L^2: y_c = Fw_c,~w_c\in L^2\right\}.
\]
It is proved in \cite{NagOguYam2011} that
the band-limited signal subspace
\[
 BL := \left\{y_c\in L^2: \text{supp~} \hat{y}_c \subset \left(-\pi/h,\pi/h\right)\right\},
\]
is a proper subset of $FL^2$,
that is, $BL \subsetneq FL^2$.
The filter $F(s)$ is chosen based on the following guidelines:
\begin{itemize}
\item a frequency distribution of input analog signals
obtained by averaging or enclosing gains of their Fourier transforms.
\item a dynamical model of signal generator such as musical instruments.
\end{itemize}
Example \ref{Ex:Johnston-yy} in Section \ref{Sec:FSFH} gives a 
brief guideline on how to choose $F(s)$ based on the 
envelope of energy distributions of the signal.  
Note that when $F$ is ideal, 
then the class we are dealing with agrees with the ideal sampling theorem.

The produced signal $y_{c}$ is then sampled 
by the ideal sampler ${\cal S}_{h}$ and becomes a discrete-time signal
$y_{d}$ with sampling period $h$. 
This signal is then upsampled by $\us L$ to 
allow for processing (interpolation) 
between the original sampling period $h$.  
The digital filter $K(z)$ 
processes this upsampled 
signal to produce $y_{K,d}$.  The signal $y_{K,d}$ then 
goes through the zero-order hold $\hold{h/L}$ 
and becomes a continuous-time signal.  
It is then further processed by an analog low-pass filter $P(s)$ 
to become the final analog output $y_{p}$.  

In the upper part of the diagram, we allow $m$ steps of 
delay for the analog signal $y_{c}$ and obtain 
$y_{c}(t - mh)$.  This is a setup for allowing a 
``preview'' of $y_c$ for $m$ samples
by the proper filter transfer function
$K(z)$.
It is very effective compared to
reconstruction without a preview.  This also takes care of
certain processing delays caused by the processing filter.  
The integer $m$ is a design parameter that can be chosen 
by the designer.  This is in marked contrast to the conventional
design methodologies:  These methods usually allow a non-causal 
impulse response for reconstruction, 
e.g., \cite{Unser05,MeinsmaMirkinSP10a,MeinsmaMirkinSP10b}.  
But in real implementation, one has to truncate it, and 
it is often unclear how many steps one would need to obtain 
a desired accuracy.  In the present setup, one can {\em prespecify
an allowable step of delays\/} (preview), and obtain an optimal design 
under such a constraint.  

Finally, the processed signal 
$y_{p}$ is compared with this delayed $y_{c}(t - mh)$ 
and subtracted from it to obtain the error signal 
$e_{c}$.  The design objective is to make the 
error as small as possible.  
Observe also that this design framework is 
formulated in the {\em continuous-time domain\/} in contrast to 
the usual discrete-time setups.  

We must specify a performance index to give a precise meaning
to this problem.  The following $L^{2}$ induced 
norm from $w_{c}$ to $e_{c}$ 
(or the sampled-data $H^{\infty}$ norm) is the one 
we take: 
\begin{equation} \label{eqn:perfindex0}
  J := \sup_{w_{c}\in L^{2}, w_{c}\neq 0} 
             \frac{\Twonorm{e_{c}}}{\Twonorm{w_{c}}}.
\end{equation}
We thus arrive at the following design problem: 

\begin{problem}
\label{prob:Hinf} 
Let $T_{ew}$ denote the input/output operator 
from $w_c$ to $e_c(\cdot):=y_c(\cdot)-u_c(\cdot-mh)$ 
in Fig.~\ref{Fig:err_interp}.  
Given an attenuation level $\gamma>0$, find, if one exists, 
a stable digital (discrete-time) filter $K(z)$ such that
\begin{equation} \label{eqn:perfindex1}
	\Hinfnorm{T_{ew}}:= J = \sup_{w_c \in L^2[0,\infty)}
	    \frac{\Twonorm{T_{ew}w_c}}{\Twonorm{w_c}} < \gamma .
\end{equation}
\end{problem}

The performance index (\ref{eqn:perfindex1}) intends to 
minimize the maximum error induced by an (unknown) input 
$w_{c}$ that gives rise to the largest norm of $e_{c}$ 
among all inputs.  This is made possible by the 
$H^{\infty}$ design methodology.  Note that the actual error 
is not known, but due to the min-max nature of the problem,
we can minimize the worst transmission error.  Observe also
that this setup allows for a capability of minimizing 
continuous-time phase errors due to the continuous-time nature
of the performance index, as opposed to the conventional 
gain-phase design principles.  

This min-max problem differs sharply from the orthogonal 
projection based methods.  Also, due to sampling, $T_{ew}$ is 
not even time-invariant (in continuous-time).

It is now known however that this problem is reducible 
to a linear time-invariant problem 
via {\em lifting}; see Appendix \ref{Sec:appendLifting};
the problem is now solvable 
via now-standard $\Hinf$ control theory, see, e.g., 
\cite{ChenFrancisBook,BamiehPearson92} (see also \cite{DGKF} for standard 
treatments of $H^{\infty}$ control in the continuous-time setting).  

The existence of $e^{-mhs}$ makes this an infinite-dimensional 
$\Hinf$ problem; see \cite{PPKYYCDC95,%
YYAutoma99,YYCDC02,YYMTNS2000}, etc.  
The simplest one is to employ the so-called fast-sampling/fast-hold
approximation, which we will outline in the next section.

\section{Solution via Fast Sample/Hold}
\label{Sec:FSFH}

While Problem \ref{prob:Hinf} is known to be 
reducible to a finite-dimensional problem \cite{PPKYYCDC95,Mirkin}, 
it is not necessarily appealing computationally.
It is often more convenient to resort 
to an approximation method.  We employ the fast sample/hold 
approximation \cite{KellerAndersonTAC92,ChenFrancisBook,YYAutoma99,YYCDC02}. 
This method approximates 
continuous-time inputs and outputs via a sampler and hold
that operate in the period $h/N$ for some positive integer $N$.
The method usually works fairly well for $N \sim 5L$, where
$L$ is the upsampling ratio given in Section \ref{Sec:SignalReconstruction}, 
and the convergence of such 
an approximation is shown in \cite{YYAutoma99,YYCDC02}.
We show here the design procedure of $K(z)$ by the fast
sampling/hold approximation.

\begin{figure}[tbp]
\centering
\unitlength 0.1in
\begin{picture}( 45.0000, 14.4000)(  2.0000,-16.4000)
%
\special{pn 8}%
\special{pa 200 920}%
\special{pa 560 920}%
\special{fp}%
\special{sh 1}%
\special{pa 560 920}%
\special{pa 494 900}%
\special{pa 508 920}%
\special{pa 494 940}%
\special{pa 560 920}%
\special{fp}%
%
\special{pn 8}%
\special{pa 560 680}%
\special{pa 1040 680}%
\special{pa 1040 1160}%
\special{pa 560 1160}%
\special{pa 560 680}%
\special{fp}%
%
\special{pn 8}%
\special{pa 1040 920}%
\special{pa 1160 920}%
\special{fp}%
%
\special{pn 8}%
\special{pa 1160 440}%
\special{pa 1160 1400}%
\special{fp}%
%
\special{pn 8}%
\special{pa 1160 1400}%
\special{pa 1400 1400}%
\special{fp}%
\special{sh 1}%
\special{pa 1400 1400}%
\special{pa 1334 1380}%
\special{pa 1348 1400}%
\special{pa 1334 1420}%
\special{pa 1400 1400}%
\special{fp}%
%
\special{pn 8}%
\special{pa 1400 1160}%
\special{pa 1880 1160}%
\special{pa 1880 1640}%
\special{pa 1400 1640}%
\special{pa 1400 1160}%
\special{fp}%
%
\special{pn 8}%
\special{pa 3560 1160}%
\special{pa 4040 1160}%
\special{pa 4040 1640}%
\special{pa 3560 1640}%
\special{pa 3560 1160}%
\special{fp}%
%
\special{pn 8}%
\special{pa 4040 1400}%
\special{pa 4280 1400}%
\special{fp}%
%
\special{pn 8}%
\special{pa 4280 1400}%
\special{pa 4280 980}%
\special{fp}%
\special{sh 1}%
\special{pa 4280 980}%
\special{pa 4260 1048}%
\special{pa 4280 1034}%
\special{pa 4300 1048}%
\special{pa 4280 980}%
\special{fp}%
%
\special{pn 8}%
\special{ar 4280 920 60 60  0.0000000 6.2831853}%
%
\special{pn 8}%
\special{pa 2360 200}%
\special{pa 3080 200}%
\special{pa 3080 680}%
\special{pa 2360 680}%
\special{pa 2360 200}%
\special{fp}%
%
\special{pn 8}%
\special{pa 4280 440}%
\special{pa 4280 860}%
\special{fp}%
\special{sh 1}%
\special{pa 4280 860}%
\special{pa 4300 794}%
\special{pa 4280 808}%
\special{pa 4260 794}%
\special{pa 4280 860}%
\special{fp}%
%
\special{pn 8}%
\special{pa 4340 920}%
\special{pa 4700 920}%
\special{fp}%
\special{sh 1}%
\special{pa 4700 920}%
\special{pa 4634 900}%
\special{pa 4648 920}%
\special{pa 4634 940}%
\special{pa 4700 920}%
\special{fp}%
\put(27.2000,-4.4000){\makebox(0,0){$e^{-mhs}$}}%
\put(8.0000,-9.2000){\makebox(0,0){$F(s)$}}%
\put(16.4000,-14.0000){\makebox(0,0){$\samp{h}$}}%
\put(38.0000,-14.0000){\makebox(0,0){$P(s)$}}%
\put(2.0000,-8.6000){\makebox(0,0)[lb]{$w_c$}}%
\put(13.4000,-9.2000){\makebox(0,0){$y_c$}}%
\put(43.4000,-14.0000){\makebox(0,0)[lb]{$y_p$}}%
\put(43.6400,-8.6000){\makebox(0,0)[lb]{$e_c$}}%
\put(43.4000,-9.8000){\makebox(0,0)[lt]{$-$}}%
%
\special{pn 8}%
\special{pa 1880 1400}%
\special{pa 2120 1400}%
\special{fp}%
\special{sh 1}%
\special{pa 2120 1400}%
\special{pa 2054 1380}%
\special{pa 2068 1400}%
\special{pa 2054 1420}%
\special{pa 2120 1400}%
\special{fp}%
%
\special{pn 8}%
\special{pa 2120 1160}%
\special{pa 2600 1160}%
\special{pa 2600 1640}%
\special{pa 2120 1640}%
\special{pa 2120 1160}%
\special{fp}%
%
\special{pn 8}%
\special{pa 2600 1400}%
\special{pa 2840 1400}%
\special{fp}%
\special{sh 1}%
\special{pa 2840 1400}%
\special{pa 2774 1380}%
\special{pa 2788 1400}%
\special{pa 2774 1420}%
\special{pa 2840 1400}%
\special{fp}%
%
\special{pn 8}%
\special{pa 2840 1160}%
\special{pa 3320 1160}%
\special{pa 3320 1640}%
\special{pa 2840 1640}%
\special{pa 2840 1160}%
\special{fp}%
%
\special{pn 8}%
\special{pa 3320 1400}%
\special{pa 3560 1400}%
\special{fp}%
\special{sh 1}%
\special{pa 3560 1400}%
\special{pa 3494 1380}%
\special{pa 3508 1400}%
\special{pa 3494 1420}%
\special{pa 3560 1400}%
\special{fp}%
%
\special{pn 8}%
\special{pa 1160 440}%
\special{pa 2360 440}%
\special{fp}%
\special{sh 1}%
\special{pa 2360 440}%
\special{pa 2294 420}%
\special{pa 2308 440}%
\special{pa 2294 460}%
\special{pa 2360 440}%
\special{fp}%
%
\special{pn 8}%
\special{pa 3080 440}%
\special{pa 4280 440}%
\special{fp}%
\put(23.6000,-14.0000){\makebox(0,0){$\widetilde{K}\!(z)$}}%
\put(30.8000,-14.0000){\makebox(0,0){$\ghold{h}$}}%
\end{picture}%
\caption{Reduced single-rate error system $T_{ew}$}
\label{fig:urDA}
\end{figure}
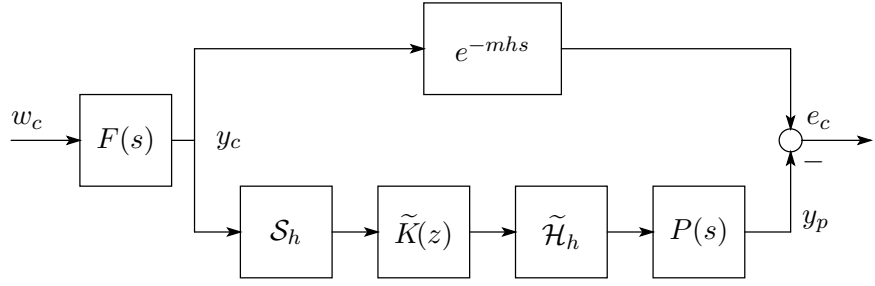

The error system in Fig.~\ref{Fig:err_interp} is a
multirate system due to the upsampler $\us{L}$.
We first reduce this system to a single-rate one.
Introduce the {\it discrete-time lifting},
also known as the {\it polyphase decomposition} \cite{Vaidyanathan},
$\dlift{L}$ and its inverse
$\idlift{L}$ as 
\begin{equation}
    \dlift{L} := (\ds{L})
	\left[\begin{array}{cccc}
	  1 & z & \cdots & z^{L-1}
	\end{array}\right]^T,\quad
    \idlift{L} :=
 	\left[\begin{array}{cccc}
	  1 & z^{-1} & \cdots & z^{-L+1}
	\end{array}\right](\us{L}).
\label{eq:dlift}
\end{equation}
Then $K(z)(\us{L})$ can be rewritten by
a lifted system as
\[
    K(z)(\us{L})
	    = \idlift{L}\widetilde{K}(z),\quad
    \widetilde{K}(z)
	    := \dlift{L}K(z)\idlift{L}
		  \left[\begin{array}{cccc}
			1&0&\cdots&0
		  \end{array}\right]^T.
\]
The filter $\widetilde{K}(z)$ is an LTI (linear and time-invariant), 
single-input/$L$-output system that satisfies
\begin{equation}
    K(z) = \left[\begin{array}{cccc}
		 1&z^{-1}&\cdots &z^{-L+1}
	   \end{array}\right] \widetilde{K}(z^L).
    \label{eq:Kz}
\end{equation}
Define $\ghold{h}:=\hold{h/L}\idlift{L}$,
and we obtain the following equality
\[
    \hold{h/L}K(z)(\us{L})\samp{h}=\ghold{h}\widetilde{K}(z)\samp{h}.
\]
Hence the multirate system in Fig.~\ref{Fig:err_interp} is reduced to 
the single-rate system shown in Fig.~\ref{fig:urDA}.

\begin{figure}[tbp]
\centering
\unitlength 0.1in
\begin{picture}( 38.4000,  4.8000)(  4.0000, -7.0100)
%
\special{pn 8}%
\special{pa 400 462}%
\special{pa 880 462}%
\special{fp}%
\special{sh 1}%
\special{pa 880 462}%
\special{pa 814 442}%
\special{pa 828 462}%
\special{pa 814 482}%
\special{pa 880 462}%
\special{fp}%
%
\special{pn 8}%
\special{pa 880 222}%
\special{pa 1600 222}%
\special{pa 1600 702}%
\special{pa 880 702}%
\special{pa 880 222}%
\special{fp}%
%
\special{pn 8}%
\special{pa 1600 462}%
\special{pa 2080 462}%
\special{fp}%
\special{sh 1}%
\special{pa 2080 462}%
\special{pa 2014 442}%
\special{pa 2028 462}%
\special{pa 2014 482}%
\special{pa 2080 462}%
\special{fp}%
%
\special{pn 8}%
\special{pa 2080 222}%
\special{pa 2560 222}%
\special{pa 2560 702}%
\special{pa 2080 702}%
\special{pa 2080 222}%
\special{fp}%
%
\special{pn 8}%
\special{pa 2560 462}%
\special{pa 3040 462}%
\special{fp}%
\special{sh 1}%
\special{pa 3040 462}%
\special{pa 2974 442}%
\special{pa 2988 462}%
\special{pa 2974 482}%
\special{pa 3040 462}%
\special{fp}%
%
\special{pn 8}%
\special{pa 3040 222}%
\special{pa 3760 222}%
\special{pa 3760 702}%
\special{pa 3040 702}%
\special{pa 3040 222}%
\special{fp}%
%
\special{pn 8}%
\special{pa 3760 462}%
\special{pa 4240 462}%
\special{fp}%
\special{sh 1}%
\special{pa 4240 462}%
\special{pa 4174 442}%
\special{pa 4188 462}%
\special{pa 4174 482}%
\special{pa 4240 462}%
\special{fp}%
\put(23.2000,-4.6100){\makebox(0,0){$T_{ew}$}}%
\put(12.4000,-4.6100){\makebox(0,0){$\hold{h/N}$}}%
\put(34.0000,-4.6100){\makebox(0,0){$\samp{h/N}$}}%
\put(4.0000,-4.0100){\makebox(0,0)[lb]{$\tilde{w}_d$}}%
\put(16.6000,-4.0100){\makebox(0,0)[lb]{$w_c$}}%
\put(26.2000,-4.0100){\makebox(0,0)[lb]{$e_c$}}%
\put(40.6000,-4.0100){\makebox(0,0)[lb]{$\tilde{e}_d$}}%
\end{picture}%
\caption{Fast discretization}
\label{fig:fsfh_fig}
\end{figure}
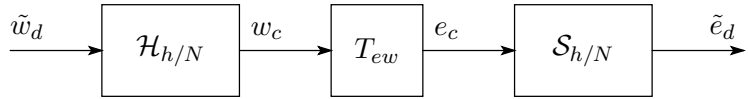

We then employ the fast sample/hold approximation for
the error system $T_{ew}$ in Fig.~\ref{fig:urDA}.
We connect fast sample and hold devices $\samp{h/N}$, $\hold{h/N}$
with the error system $T_{ew}$ as shown in Fig.~\ref{fig:fsfh_fig}.
For brevity of notation, let us adopt the following 
shorthand notation for the transfer function 
$D + C(zI - A)^{-1}B$: 
\begin{equation} \label{eqn:Doylenotation}
  \ssr{A}{B}{C}{D} := D + C(zI - A)^{-1}B 
\end{equation}

The sampled-data error system $T_{ew}$ can be approximated by 
a discrete-time LTI system as in the following theorem:
\begin{theorem}
\label{th:fsfh}
Let state-space realizations of $F(s)$ and $P(s)$ be given by 
\[
 F(s) = \ssr{A_{Fc}}{B_{Fc}}{C_F}{0},\quad  P(s) = \ssr{A_{Pc}}{B_{Pc}}{C_P}{D_P}.
\]
Let $N=Ll$ where $l$ is a positive integer, and 
define the discrete-time LTI system $T_{N}$ as follows: 
\begin{equation}
  T_N(z) = z^{-m}F_N(z)-P_N(z)H\widetilde{K}(z)SF_N(z),
  \label{eq:fsfh}
\end{equation}
{\small
\[
 \begin{split}
  F_N &=
    \left[\begin{array}{c|cccc}
	 A_F^N & A_F^{N\!-\!1}B_F & A_F^{N-2}B_F & \ldots & B_F\\\hline
	 C_F & 0 & 0 & \ldots & 0\\
	 C_FA_F & C_FB_F & 0 & \ddots & \vdots\\
	 \vdots & \vdots & \vdots & \ddots & 0\\
	 C_FA_F^{N-1}\! & \!C_FA_F^{N-2}B_F \!&\! C_FA_F^{N-3}B_F \!& \!\ldots \!& \!0
    \end{array}\right],\\
  P_N &=
    \left[\begin{array}{c|cccc}
	 A_P^N & A_P^{N\!-\!1}B_P & A_P^{N-2}B_P & \ldots & B_P\\\hline
	 C_P & D_P & 0 & \ldots & 0\\
	 C_PA_P & C_PB_P & D_P & \ddots & \vdots\\
	 \vdots & \vdots & \vdots & \ddots & 0\\
	 C_PA_P^{N-1}\! & \!C_PA_P^{N-2}B_P \!&\! C_PA_P^{N-3}B_P \!& \!\ldots \!& \!D_P
    \end{array}\right],
 \end{split}
\]}
\begin{gather*}
  A_F = e^{A_{Fc}h/N},\quad B_F = \int_0^{h/N} e^{A_{Fc}t}B_{Fc}dt,\quad
  A_P = e^{A_{Pc}h/N},\quad B_P = \int_0^{h/N} e^{A_{Pc}t}B_{Pc}dt,\\
  H := \diag{\{I_l\}}\in\R^{N\times L},\quad I_l:=[1,1,\ldots ,1]^T\in\R^{l},\quad
  S := [1,0,\ldots,0] \in \R^{1\times N}.
\end{gather*}
Then, for each fixed $\widetilde{K}$ and for 
each $\omega \in [0, 2\pi/h)$, the frequency response 
\begin{equation}
  \|T_{N}(e^{j\omega h})\| \rightarrow
       \|T_{ew}(e^{j\omega h})\|.
 \label{eq:fsfh-convergence}
\end{equation}
as $N\rightarrow \infty$, and this convergence is 
uniform with respect to $\omega \in [0, 2\pi/h)$.  
Furthermore, this convergence is also uniform 
in $\widetilde{K}$ if $\widetilde{K}$ ranges over 
a compact set of filters.  
\end{theorem}

\Proof
See Appendix \ref{Sec:appendFSFH}.
\EndProof

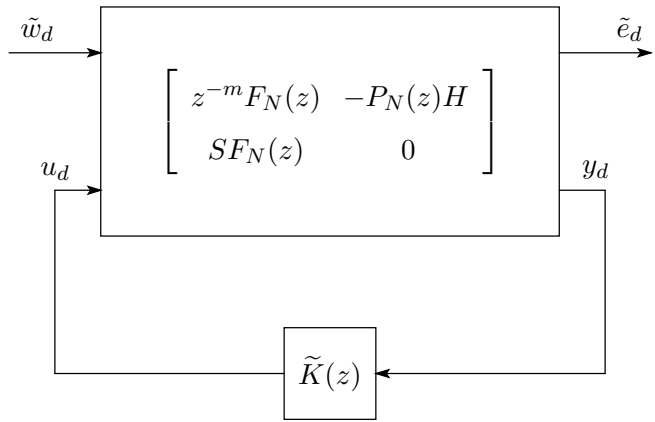
\begin{figure}[tbp]
\centering
\unitlength 0.1in
\begin{picture}( 33.6000, 21.6000)(  4.0000,-23.8100)
%
\special{pn 8}%
\special{pa 400 462}%
\special{pa 880 462}%
\special{fp}%
\special{sh 1}%
\special{pa 880 462}%
\special{pa 814 442}%
\special{pa 828 462}%
\special{pa 814 482}%
\special{pa 880 462}%
\special{fp}%
%
\special{pn 8}%
\special{pa 880 222}%
\special{pa 3280 222}%
\special{pa 3280 1422}%
\special{pa 880 1422}%
\special{pa 880 222}%
\special{fp}%
%
\special{pn 8}%
\special{pa 3280 462}%
\special{pa 3760 462}%
\special{fp}%
\special{sh 1}%
\special{pa 3760 462}%
\special{pa 3694 442}%
\special{pa 3708 462}%
\special{pa 3694 482}%
\special{pa 3760 462}%
\special{fp}%
%
\special{pn 8}%
\special{pa 640 1182}%
\special{pa 880 1182}%
\special{fp}%
\special{sh 1}%
\special{pa 880 1182}%
\special{pa 814 1162}%
\special{pa 828 1182}%
\special{pa 814 1202}%
\special{pa 880 1182}%
\special{fp}%
%
\special{pn 8}%
\special{pa 640 1182}%
\special{pa 640 2142}%
\special{fp}%
%
\special{pn 8}%
\special{pa 640 2142}%
\special{pa 1840 2142}%
\special{fp}%
%
\special{pn 8}%
\special{pa 1840 1902}%
\special{pa 2320 1902}%
\special{pa 2320 2382}%
\special{pa 1840 2382}%
\special{pa 1840 1902}%
\special{fp}%
%
\special{pn 8}%
\special{pa 3280 1182}%
\special{pa 3520 1182}%
\special{fp}%
%
\special{pn 8}%
\special{pa 3520 1182}%
\special{pa 3520 2142}%
\special{fp}%
%
\special{pn 8}%
\special{pa 3520 2142}%
\special{pa 2320 2142}%
\special{fp}%
\special{sh 1}%
\special{pa 2320 2142}%
\special{pa 2388 2162}%
\special{pa 2374 2142}%
\special{pa 2388 2122}%
\special{pa 2320 2142}%
\special{fp}%
\put(20.8000,-21.4100){\makebox(0,0){$\widetilde{K}(z)$}}%
\put(20.8000,-8.2100){\makebox(0,0){$\Gd$}}%
\put(4.6000,-4.0100){\makebox(0,0)[lb]{$\tilde{w}_d$}}%
\put(35.8000,-4.0100){\makebox(0,0)[lb]{$\tilde{e}_d$}}%
\put(34.0000,-11.2100){\makebox(0,0)[lb]{$y_d$}}%
\put(5.6500,-11.2700){\makebox(0,0)[lb]{$u_d$}}%
\end{picture}%
\caption{Discrete-time system for $H^\infty$ optimization}
\label{fig:gplant_fsfh}
\end{figure}

In view of the uniformity of convergence $\Hinfnorm{T_{N}}$ 
in $\widetilde{K}$, our design problem (\ref{eqn:perfindex1}) 
can be approximated by 
\begin{displaymath}
    \Hinfnorm{T_{N}} < \gamma.
\end{displaymath} 
This is a discrete-time $H^\infty$ optimization problem.
To obtain a filter $\widetilde{K}(z)$ satisfying the above inequality, 
we can adopt numerical softwares
as MATLAB with robust control toolbox
\cite{MatlabRCT},
by the generalized plant representation depicted in
Fig.~\ref{fig:gplant_fsfh},
where $\tilde{w}_d=\dlift{N}w_d$ and $\tilde{e}_d=\dlift{N}e_d$.
Once the optimal filter $\widetilde{K}(z)$ is obtained, 
one can obtain the interpolation filter $K(z)$ by 
formula (\ref{eq:Kz}).  

\begin{example} \label{Ex:Johnston-yy}
Let us make a comparison with a usual linear phase 
filter---the Johnston filter \cite{Johnston}.  
We design the proposed filter $K(z)$ with
interpolation ratio $L = 2$, sampling period $h=1$, 
and delay step 
$m=4$.
The analog filters $F(s)$ and $P(s)$ are given by
\[
 F(s) = \frac{1}{(Ts+1)(0.1Ts+1)},~T=7.0187,~P(s)=1.
\]
Reflecting a typical energy distribution of orchestral music,  
the time constant $T = 7.0187$ is taken to be equivalent to $1$ kHz with 
sampling frequency $44.1$ kHz.  It therefore corresponds to an 
energy distribution that decays by $-20$ dB per decade from 
$1$ kHz and $-40$ dB per decade from $10$ kHz.

\begin{figure}[tbp]
 \centering
 \includegraphics[width=0.48\linewidth]{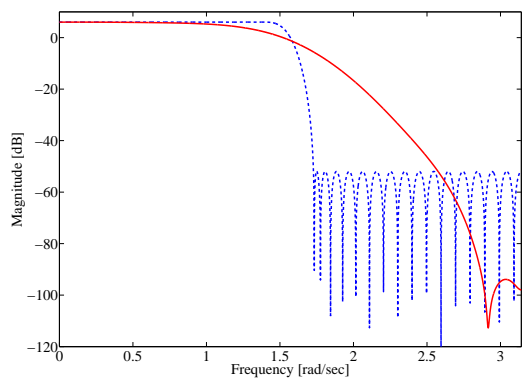}
 \caption{Bode gain plot of the proposed filter (solid) and the Johnston filter (dash)}
 \label{Fig:filters_yy_johnston}
\end{figure}
Fig.~\ref{Fig:filters_yy_johnston} shows the Bode gain plots of the proposed filter and the Johnston FIR filter 
with 32 taps.
We can see that the Johnston filter has a sharp decay around the cutoff frequency 
$\omega=\pi/2$, while the filter obtained by the proposed method shows a rather mild decay.

\begin{figure}[tbp]
 \begin{tabular} {cc}
  \begin{minipage} {0.48\hsize}
   \includegraphics[width=\linewidth]{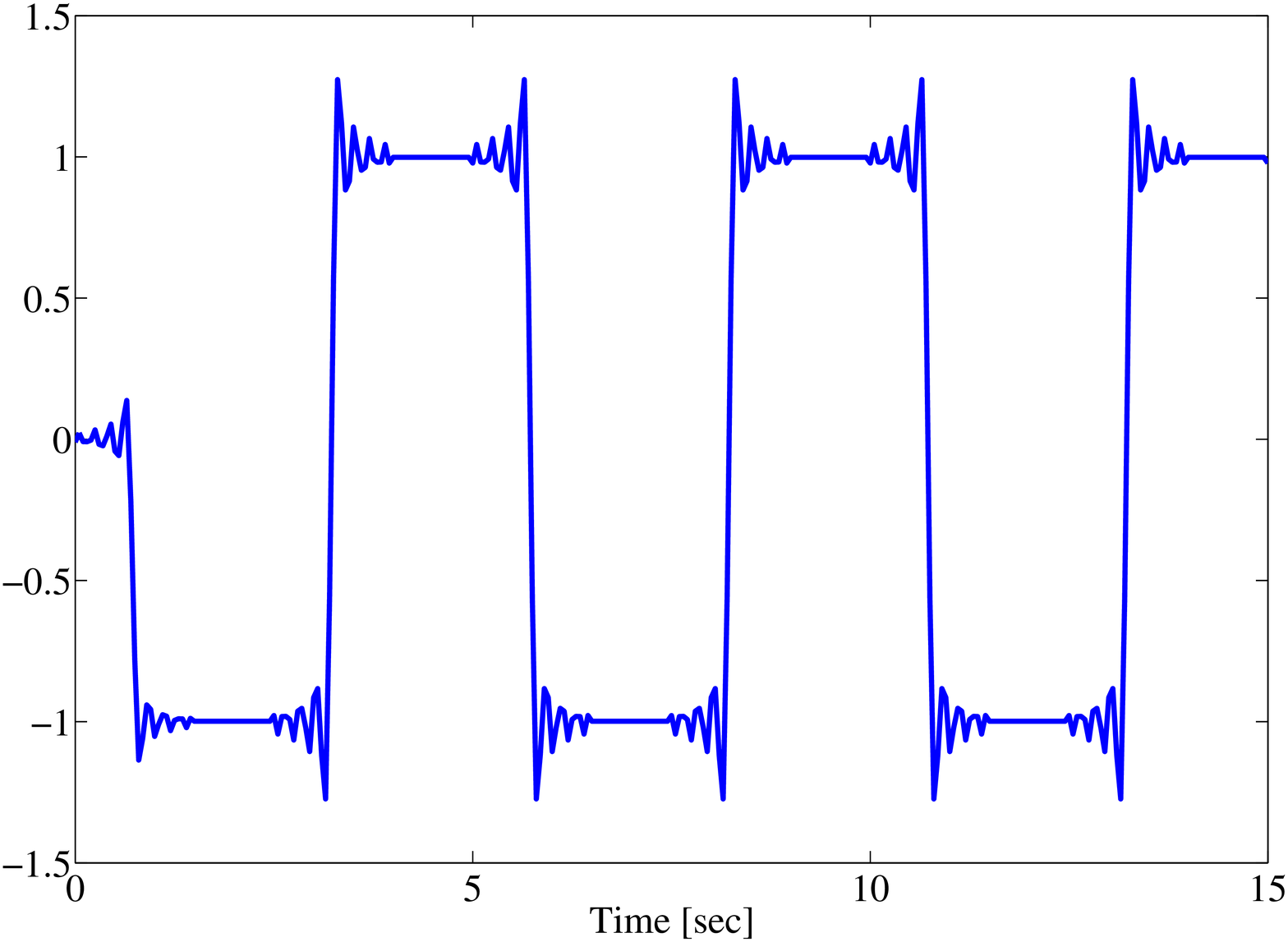}
   \caption{Response of the Johnston filter against a rectangular wave}
   \label{Fig:ringing}
  \end{minipage}
  \begin{minipage} {0.48\hsize}
   \includegraphics[width=\linewidth]{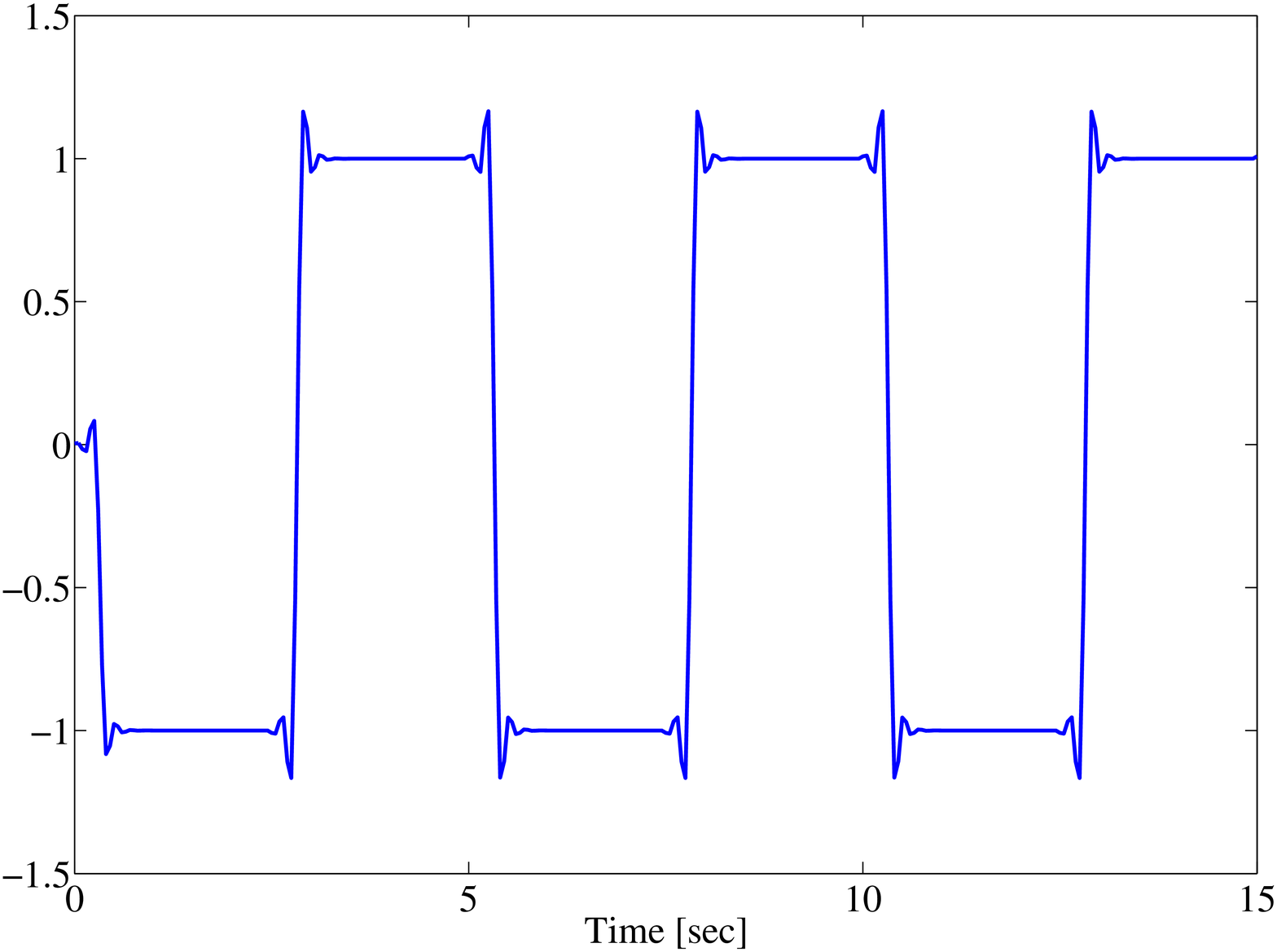}
   \caption{Response of a sampled-data design filter against a rectangular wave} 
   \label{Fig:square_wavresp}
  \end{minipage}
 \end{tabular}
\end{figure}
Fig.~\ref{Fig:ringing} shows the response of the Johnston 
filter against a rectangular wave.  It exhibits a very sharp 
ringing effect.  This is because the filter has a sharp cut-off
characteristic, and inevitably introduces the well-known 
Gibbs phenomenon due to the fact that the frequency components
beyond the pass-band are sharply truncated.  
In contrast, Fig.~\ref{Fig:square_wavresp} shows the response of the
filter designed by the present method.  
It shows virtually no ringing.
To see the difference more precisely,
we give the reconstruction error plots by the conventional method
and the proposed one for the case of $M=2$ 
in Fig.~\ref{Fig:errors}.
\begin{figure}[tbp]
 \centering
  \includegraphics[width=0.48\linewidth]{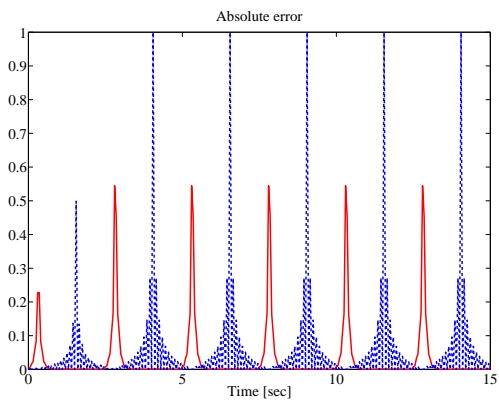}
 \caption{Absolute values of reconstruction errors: proposed (solid) and conventional (dash)}
 \label{Fig:errors}
\end{figure}
Clearly, our method offers much better filter performance.
Fig.~\ref{Fig:freqrespM2} shows the frequency response
of the sampled-data error system $T_{ew}$.
The Johnston filter exhibits large errors in the whole frequency range.
These errors give an explanation of the ringing effect in Fig.~\ref{Fig:ringing}.

\begin{figure}[tbp]
 \centering
  \includegraphics[width=0.48\linewidth]{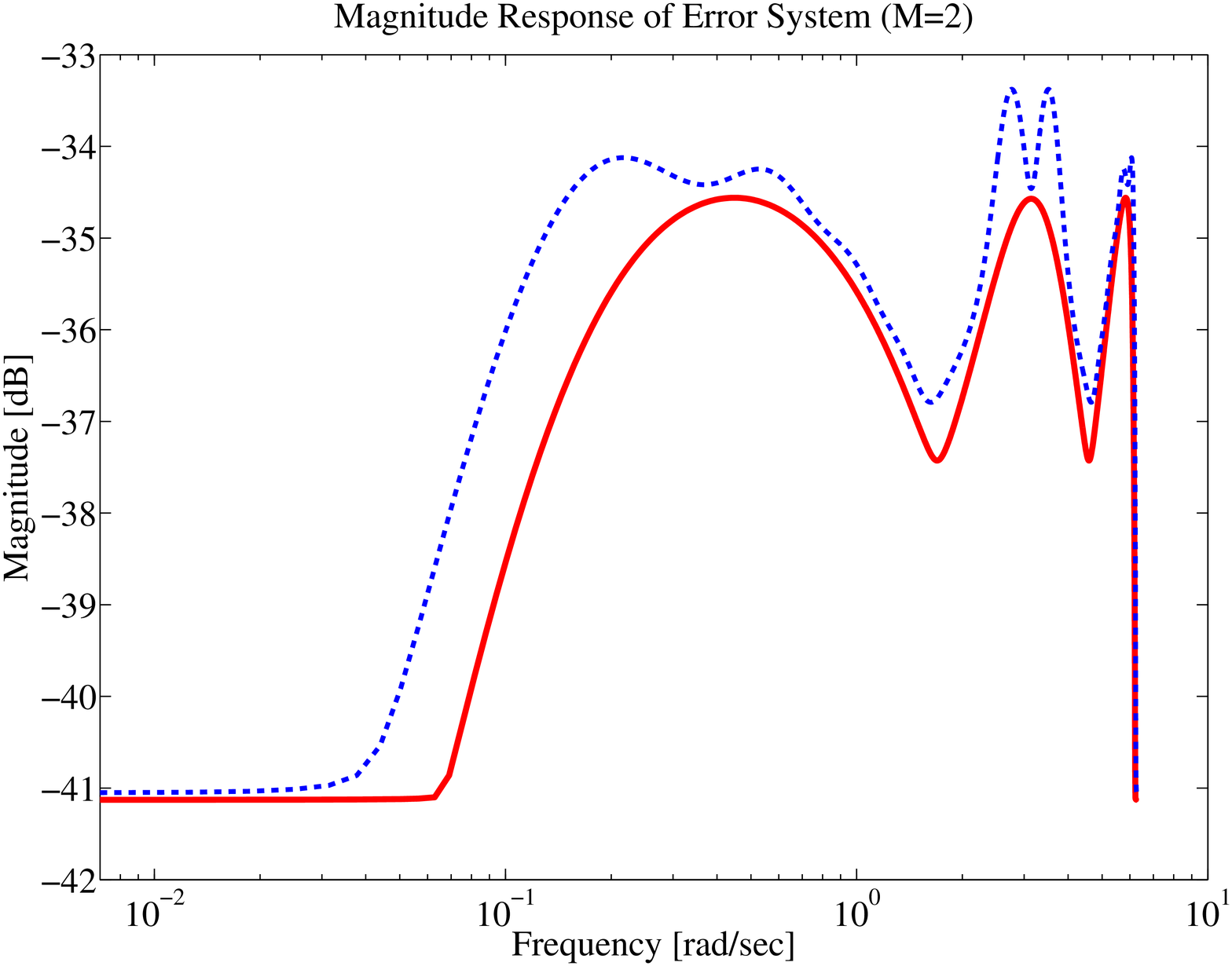}
 \caption{Frequency response of error system $T_{ew}$ with sampled-data designed filter (solid) 
   and the Johnston filter (dash)}
 \label{Fig:freqrespM2}
\end{figure}
\end{example}

\section{Comparison with other methods}
\label{Sec:Comparison}

\subsection{Remark on the consistency requirement}
\label{Subsec:consistency}

As noted in the Introduction, the notion of consistent 
reconstruction is quite widely accepted in the literature, 
e.g., \cite{Unser94,Unser00}.  
We start with a discussion of this property.  

Consider the spaces $V_{f}$ and $V_{\phi}$ in 
(\ref{eqn:generationspace}) and (\ref{eqn:reconstructionsp}).  
When $V_{f}$ and $V_{\phi}$ are equal, we are in the 
situation of orthogonal projection.  The Shannon 
paradigm is a particular case. 

When $V_{f} \neq V_{\phi}$, 
there is freedom in choosing an oblique projection, but 
consistency requirement \cite{Unser94} makes it unique:
that is, one takes the oblique projection of $L^{2}$ 
onto $V_{\phi}$ perpendicular to $V_{f}$.  
This gives a perfect reconstruction for elements in 
$V_{\phi}$ and also consistency.  That is, when one 
injects any reconstructed signal $\sum_{k} c[k]\phi(t - kh)$ 
into the acquisition device $F$ and sampling, 
one should obtain the same sampled data $c[k]$ \cite{Unser94,Unser00}.  
Unlike the least square error case, however, this process 
may yield a large error when the two spaces $V_{f}$ and
$V_{\phi}$ are apart.  This can be seen from an error 
analysis given in \cite{Unser94,Unser00} that depends
on the angle of two spaces $V_{f}$ and $V_{\phi}$.  

\begin{figure}[tbp]
 \begin{tabular} {cc}
  \begin{minipage} {0.48\hsize}
    \includegraphics[width=0.8\linewidth]{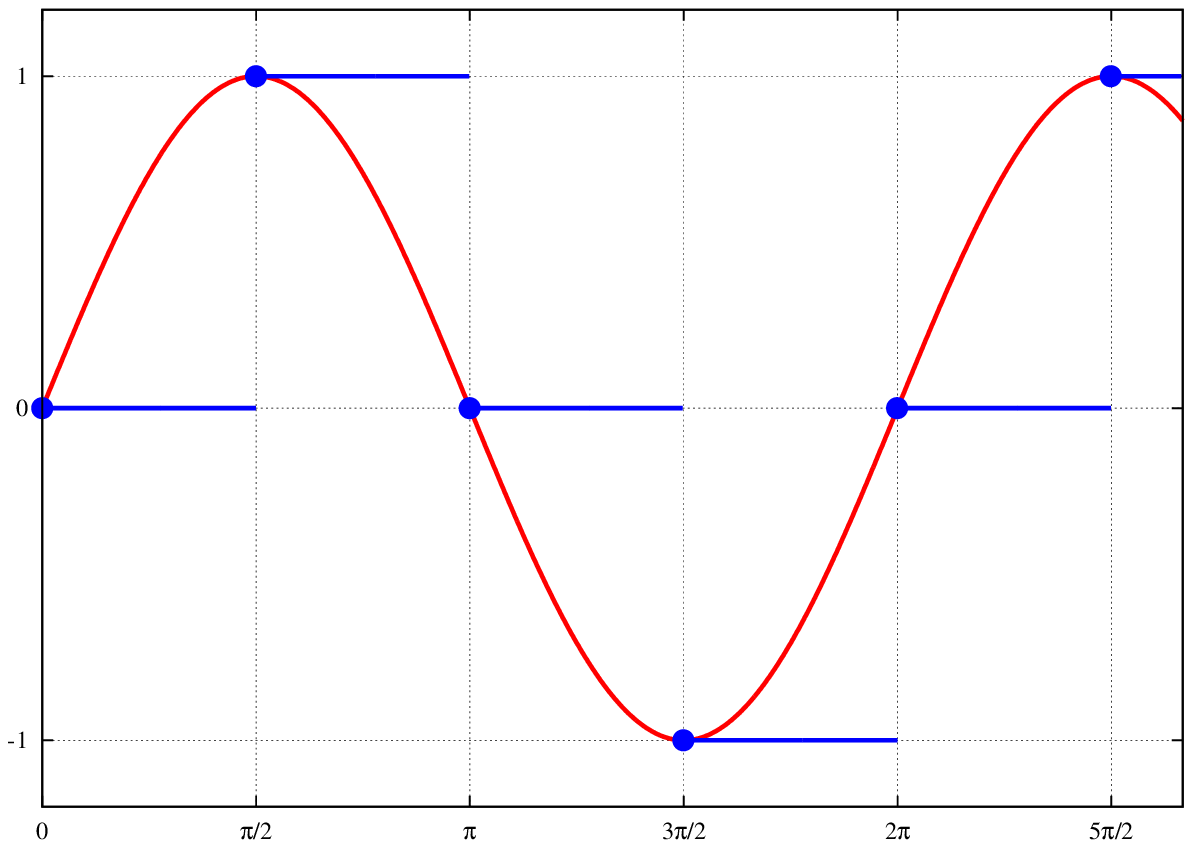}
    \caption{Consistent reconstruction \protect(\ref{eqn:consistenty})}
    \label{Fig:sinusoid1}
  \end{minipage}
  \begin{minipage} {0.48\hsize}
   \includegraphics[width=0.8\linewidth]{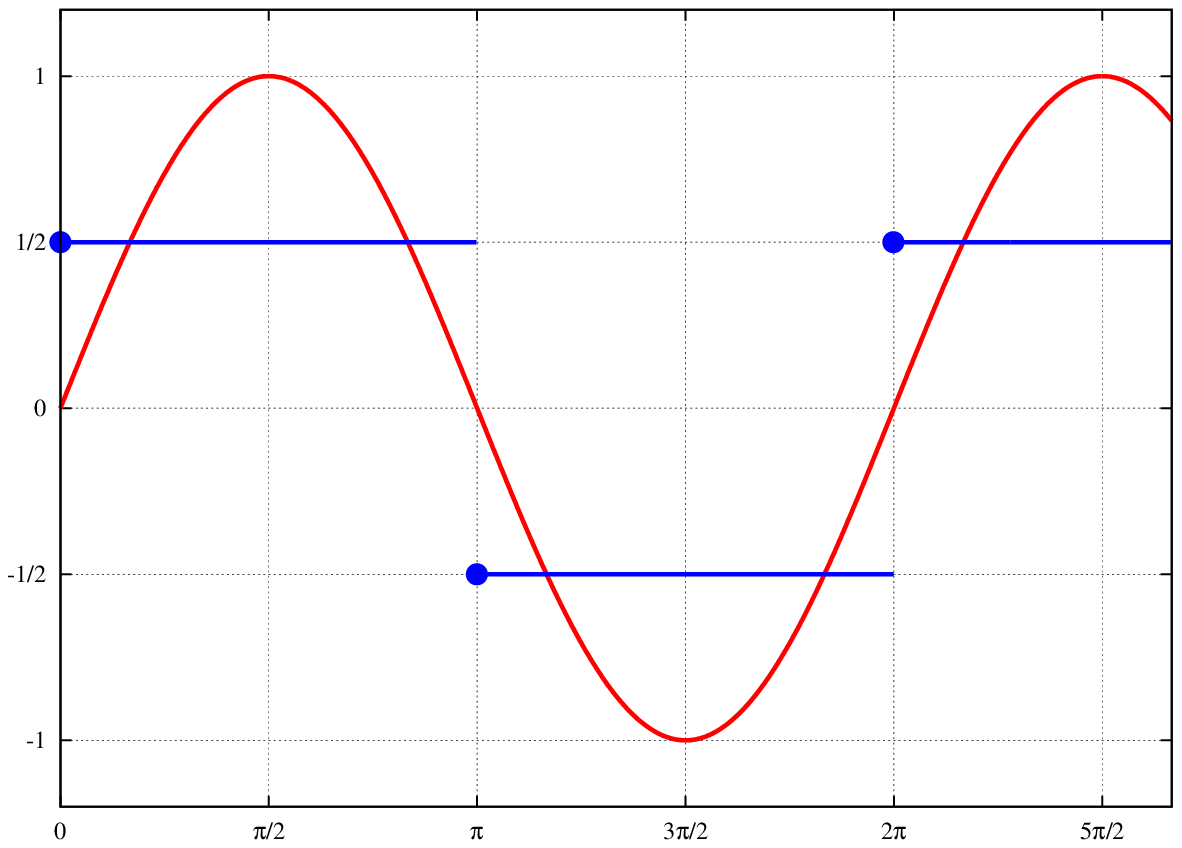}
   \caption{Mid value reconstruction \protect(\ref{eqn:midvalue})}
   \label{Fig:sinusoid2}
 \end{minipage}
\end{tabular}
\end{figure}

To see this more clearly, 
take a pure sinusoid $\sin t$ (over some bounded interval if
we strictly require it to belong to $L^{2}$), and suppose that 
we sample it with sampling period $h = \pi/2$,  
and define the acquisition device to be the ideal filter and 
the reconstruction device to be the one given by
\[
  \phi(t) = 
    \begin{cases}
       1, & 0 \leq t \leq h \\
       0, & \mbox{otherwise}.
    \end{cases} 
\]
If we sample $\sin t$ at times $2n\pi/h$, 
$n = 0, 1, 2, \ldots$, and if we adopt the consistency 
requirement (i.e., sampled $y$ gives the same values 
that we started with), the reconstructed signal 
$y$ becomes 
\begin{equation} \label{eqn:consistenty}
  y(t) = \begin{cases}
         0, & 2kh \leq t < (2k+1)h, \\
         1, & (2k+1)h \leq t < 2(k+1)h, \\
         -1, & (2k+3)h \leq t < 2(k+2)h. 
         \end{cases} 
\end{equation}
On the other hand, it is easily seen that the function
\begin{equation} \label{eqn:midvalue}
  \tilde{y}(t) = \begin{cases}
         1/2, & 2kh \leq t < 2(k+1)h, \\
         -1/2, & 2(k+1)h \leq t \leq 2(k+2)h, 
         \end{cases} 
\end{equation}
gives a better approximation for $\sin t$ with respect to the $L^{2}$ norm, 
and hence the consistency requirement does not necessarily
lead to a good approximation result.  See Figs.~\ref{Fig:sinusoid1}
and \ref{Fig:sinusoid2}.
Hence the consistent reconstruction does not necessarily 
yield a desirable result when analog performance is taken into
account.

\subsection{Comparison with cardinal exponential splines}
\label{Subsec:Unser}

Unser and Blu \cite{UnserBlu05,Unser05} 
proposed to use cardinal exponential splines 
to recover analog information from sampled data.  
The philosophy of placing emphasis on analog performance 
is exactly the same as ours here.  
Their method is however very different from the 
present one, and in some cases it does not necessarily lead to 
a desirable result.  Even a stability issue may arise.  
We here give a detailed analysis of their method, and make some comparisons.  

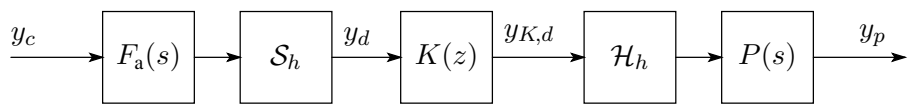
\begin{figure}[tbp]
\centering
\unitlength 0.1in
\begin{picture}( 46.8000,  4.8000)(  3.0000, -7.0100)
%
\special{pn 8}%
\special{pa 780 222}%
\special{pa 1260 222}%
\special{pa 1260 702}%
\special{pa 780 702}%
\special{pa 780 222}%
\special{fp}%
%
\special{pn 8}%
\special{pa 1260 462}%
\special{pa 1500 462}%
\special{fp}%
\special{sh 1}%
\special{pa 1500 462}%
\special{pa 1434 442}%
\special{pa 1448 462}%
\special{pa 1434 482}%
\special{pa 1500 462}%
\special{fp}%
%
\special{pn 8}%
\special{pa 1500 222}%
\special{pa 1980 222}%
\special{pa 1980 702}%
\special{pa 1500 702}%
\special{pa 1500 222}%
\special{fp}%
%
\special{pn 8}%
\special{pa 1980 462}%
\special{pa 2340 462}%
\special{fp}%
\special{sh 1}%
\special{pa 2340 462}%
\special{pa 2274 442}%
\special{pa 2288 462}%
\special{pa 2274 482}%
\special{pa 2340 462}%
\special{fp}%
%
\special{pn 8}%
\special{pa 2340 222}%
\special{pa 2820 222}%
\special{pa 2820 702}%
\special{pa 2340 702}%
\special{pa 2340 222}%
\special{fp}%
%
\special{pn 8}%
\special{pa 3300 222}%
\special{pa 3780 222}%
\special{pa 3780 702}%
\special{pa 3300 702}%
\special{pa 3300 222}%
\special{fp}%
%
\special{pn 8}%
\special{pa 3780 462}%
\special{pa 4020 462}%
\special{fp}%
\special{sh 1}%
\special{pa 4020 462}%
\special{pa 3954 442}%
\special{pa 3968 462}%
\special{pa 3954 482}%
\special{pa 4020 462}%
\special{fp}%
%
\special{pn 8}%
\special{pa 4020 222}%
\special{pa 4500 222}%
\special{pa 4500 702}%
\special{pa 4020 702}%
\special{pa 4020 222}%
\special{fp}%
%
\special{pn 8}%
\special{pa 4500 462}%
\special{pa 4980 462}%
\special{fp}%
\special{sh 1}%
\special{pa 4980 462}%
\special{pa 4914 442}%
\special{pa 4928 462}%
\special{pa 4914 482}%
\special{pa 4980 462}%
\special{fp}%
%
\special{pn 8}%
\special{pa 300 462}%
\special{pa 780 462}%
\special{fp}%
\special{sh 1}%
\special{pa 780 462}%
\special{pa 714 442}%
\special{pa 728 462}%
\special{pa 714 482}%
\special{pa 780 462}%
\special{fp}%
\put(10.2000,-4.6100){\makebox(0,0){$\Fa(s)$}}%
\put(17.4000,-4.6100){\makebox(0,0){$\samp{h}$}}%
\put(25.8000,-4.6100){\makebox(0,0){$K(z)$}}%
\put(35.4000,-4.6100){\makebox(0,0){$\hold{h}$}}%
\put(42.6000,-4.6100){\makebox(0,0){$P(s)$}}%
\put(3.0000,-4.0100){\makebox(0,0)[lb]{$y_c$}}%
\put(20.4000,-4.0100){\makebox(0,0)[lb]{$y_d$}}%
\put(28.8000,-4.0100){\makebox(0,0)[lb]{$y_{K\!,d}$}}%
\put(47.4000,-4.0100){\makebox(0,0)[lb]{$y_p$}}%
%
\special{pn 8}%
\special{pa 2820 462}%
\special{pa 3300 462}%
\special{fp}%
\special{sh 1}%
\special{pa 3300 462}%
\special{pa 3234 442}%
\special{pa 3248 462}%
\special{pa 3234 482}%
\special{pa 3300 462}%
\special{fp}%
\end{picture}%
\caption{Sampled-data signal reconstruction}
\label{Fig:block}
\end{figure}

Consider the block diagram Fig.~\ref{Fig:block}.
In this figure, $\Fa(s)$  and $P(s)$ are analog filters,
and $\samp{}$ and $\hold{}$ are, respectively, 
an ideal sampler and the zero-order hold with sampling period
$h=1$.
The filter $\Fa(s)$ represents the acquisition filter for sampling, and
$P(s)$ is a postfilter 
which smooths out the output of the zero-order hold $\hold{}$.
We assume that these analog filters are causal and LTI systems defined by
\[
 \Fa(s) = \frac{\prod_{m=1}^{m_1}(s-\gamma_{1m})}{\prod_{n=1}^{n_1}(s-\alpha_{1n})},\quad
 P(s) = \frac{\prod_{m=1}^{m_2}(s-\gamma_{2m})}{\prod_{n=1}^{n_2}(s-\alpha_{2n})},\quad
 m_i \leq n_i,\quad i=1,2,
\]
with impulse responses $\fa(t)$ and $p(t)$, respectively.  
Under these assumptions, Unser \cite{Unser05} 
derived the optimal reconstruction filter 
which achieves the consistency constraint
\[
 \langle y_c-y_p, \phi(\cdot -k) \rangle = 0,\quad k=0, \pm 1, \pm 2, \ldots,
\]
where $\phi(t)=\fa(-t)$.
This idea states that there remains no extra component
in the error $y_c - y_p$ that can be expanded with elements 
in $V_{\phi}$ (see (\ref{eqn:reconstructionsp})).  
The optimal filter is obtained by the oblique projection technique
\cite{Unser94,Unser00,UnserBlu05,Unser05} as follows: 
\begin{equation}
\label{eq:unserK}
K_{\text{op}}(z) = 
	\cfrac{\Delta_{1}(z) \Delta_{2}(z)}
	{\displaystyle \sum_{k=0}^{n_1+n_2+1}\beta(k)z^{-k}},
\end{equation}
where 
\begin{equation}
\label{eq:dlocalization}
\Delta_{i}(z) := \prod_{n=1}^{n_i}(1-e^{\alpha_{in}}z^{-1}),\quad i=1,2,
\end{equation}
and $\beta(0),\beta(1),\ldots,\beta(n_1+n_2+1)$ are 
the sampled values of $\beta(t)$ which is defined by its Laplace transform 
\begin{equation}
 \hat{\beta}(s) = \frac{1-e^{-s}}{s}\Fa(s)P(s)\prod_{i=1}^{2}\prod_{n=1}^{n_i}(1-e^{\alpha_{in}}e^{-s}).
 \label{eq:laplace_beta}
\end{equation}

The filter (\ref{eq:unserK}) is generally an IIR filter.
We can prove that this filter is given via system inversion as follows: 
\begin{theorem}
\label{th:inverse}
The optimal filter $K_{\text{op}}(z)$ in (\ref{eq:unserK}) 
can be equivalently realized as
\begin{equation}
\label{eq:qz}
K_{\text{op}}(z) = \frac{1}{H_{d}(z)},
\end{equation}
where $H_{d}(z)$ is the step-invariant equivalent 
discretization of $\Fa(s)P(s)$, 
that is, if a state-space realization of $\Fa(s)P(s)$ is given by $\{A,B,C,0\}$,
then
\[
H_{d}(z) = \mathcal{S}\Fa(s)P(s)\mathcal{H}
=\left[\begin{array}{c|c} e^{A}&\int_{0}^{1}e^{A\tau}Bd\tau\\\hline C&0\end{array}\right].
\]
\end{theorem}
\Proof 
See Appendix \ref{Sec:appendProofThm1}. 
\EndProof

While our sampled-data method always yields a stable filter, 
the above optimal filter $K_{\text{op}}$ 
derived by Unser \cite{Unser05}, according to this theorem, 
can have a pole in ${\D}_+:= \{z \in {\C} : |z|\geq 1\}$, 
when the relative degree of $\Fa(s)P(s)$ is strictly greater than $2$.
This follows from the following well-known result 
on limiting zeros \cite{Ast84}: 
\begin{fact} \it 
Let $\Sigma$ be a continuous-time, 
linear time-invariant single-input, single-output system with relative degree 
strictly greater than 2, and let $\Sigma_{h}$ be its 
step-invariant discretized system with sampling period $h$.  
Then there always exists 
an $h$ such that $\Sigma_{h}$ possesses 
a zero in ${\mathbb D}_+$. 
\end{fact} \rm

Even if the sampling time ($h = 1$ in our present normalization)
is not small relative to the time-constants of $\Fa(s)$ and 
$P(s)$, the discretized system $H_{d}(z)$ can still have 
an unstable zero, thereby yielding an 
{\em unstable pole\/} of $K_{\text op}(z)$. 
In such a case, 
Unser and Blu \cite{UnserBlu05} propose to use a {\em non-causal} filter,
folding back the anti-causal part associated with the pole in 
${\mathbb D}_+$ to the negative time axis.  
This can, in principle, lead to a very long delay for reconstruction.  

\begin{figure}[tbp]
 \begin{tabular} {cc}
  \begin{minipage} {0.48\hsize}
   \includegraphics[width=\linewidth]{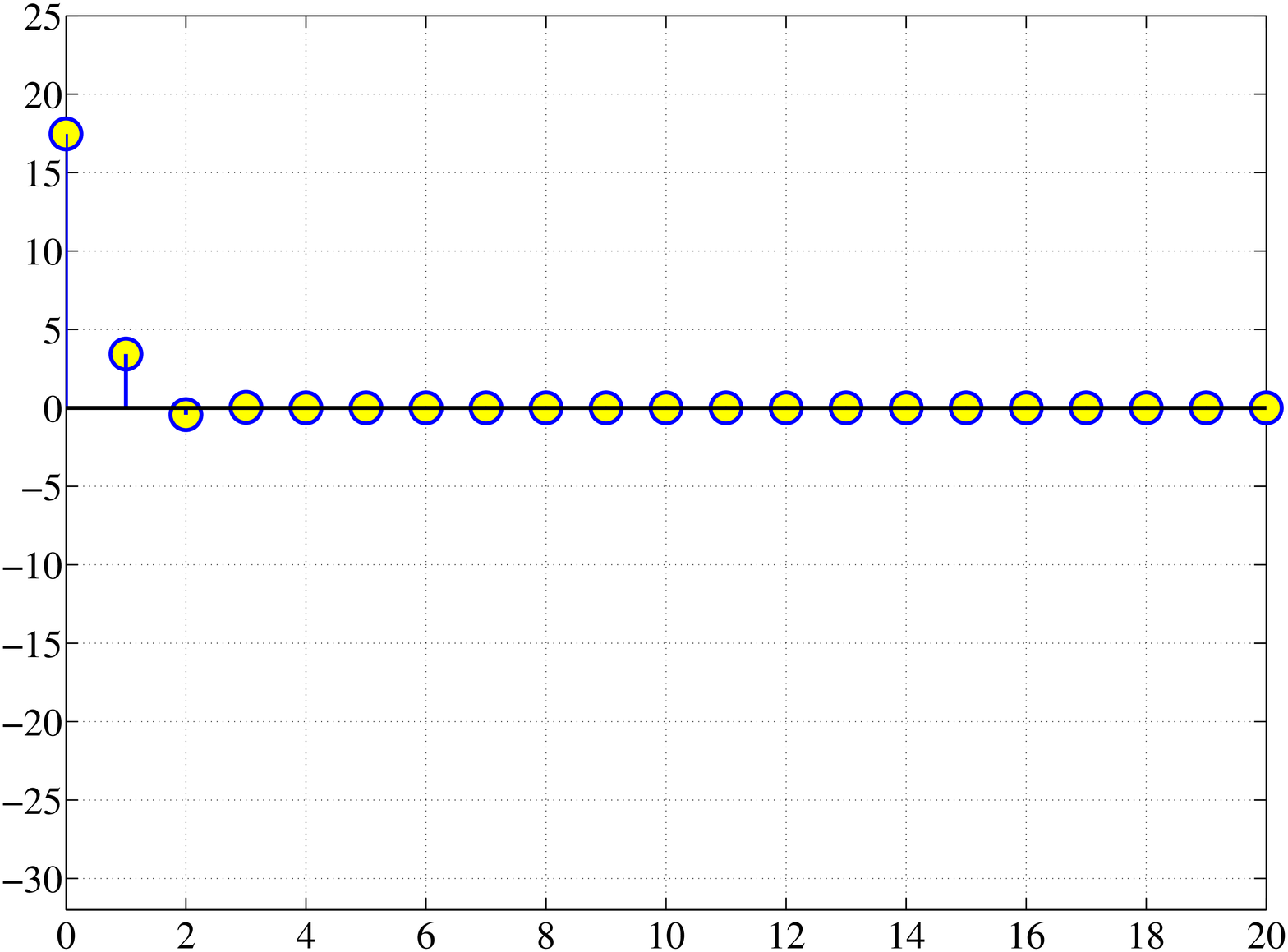}
   \caption{Impulse response of the causal part of $K_{\text{op}}(z)$}
   \label{fig:impulse_c}
  \end{minipage}
  \begin{minipage} {0.48\hsize}
   \includegraphics[width=\linewidth]{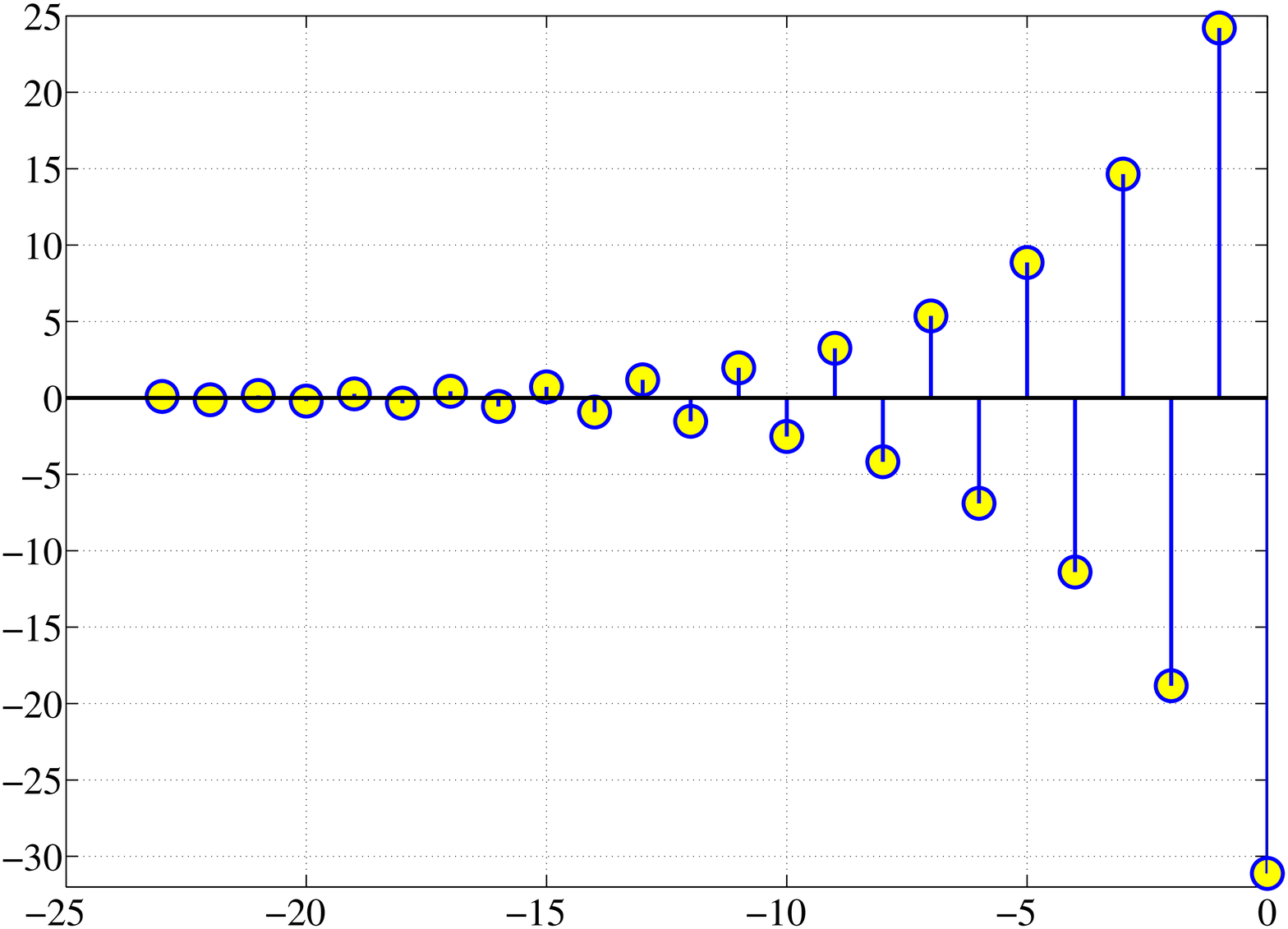}
   \caption{Impulse response of the anti-causal part of $K_{\text{op}}(z)$}
   \label{fig:impulse_ac}
  \end{minipage}
 \end{tabular}
\end{figure}

\begin{figure}[tbp]
 \begin{tabular} {cc}
  \begin{minipage} {0.48\hsize}
   \includegraphics[width=\linewidth]{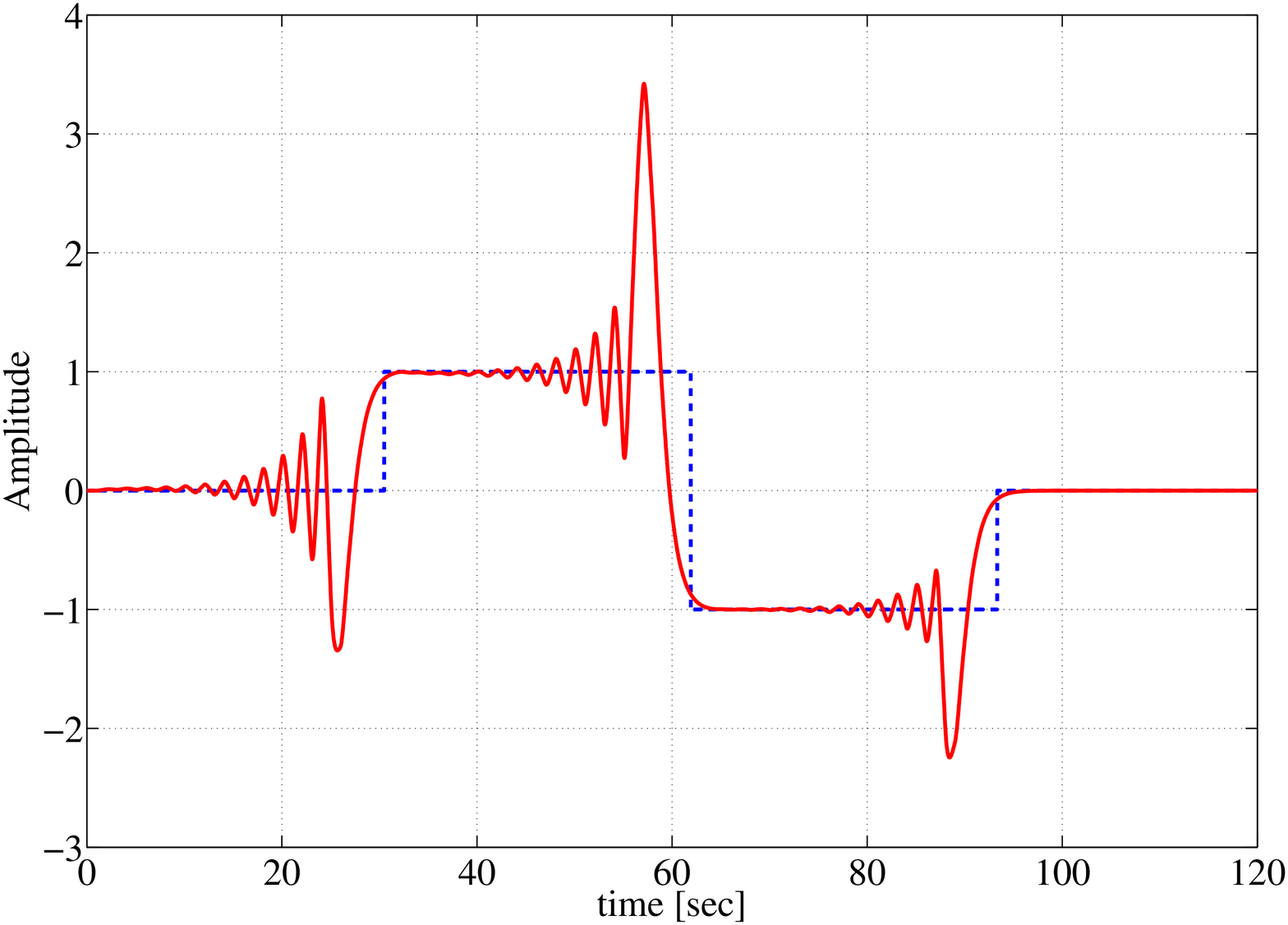}
   \caption{Reconstruction of a rectangular wave by Unser's method.}
   \label{fig:unser_reconstruction}
  \end{minipage}
  \begin{minipage} {0.48\hsize}
   \includegraphics[width=\linewidth]{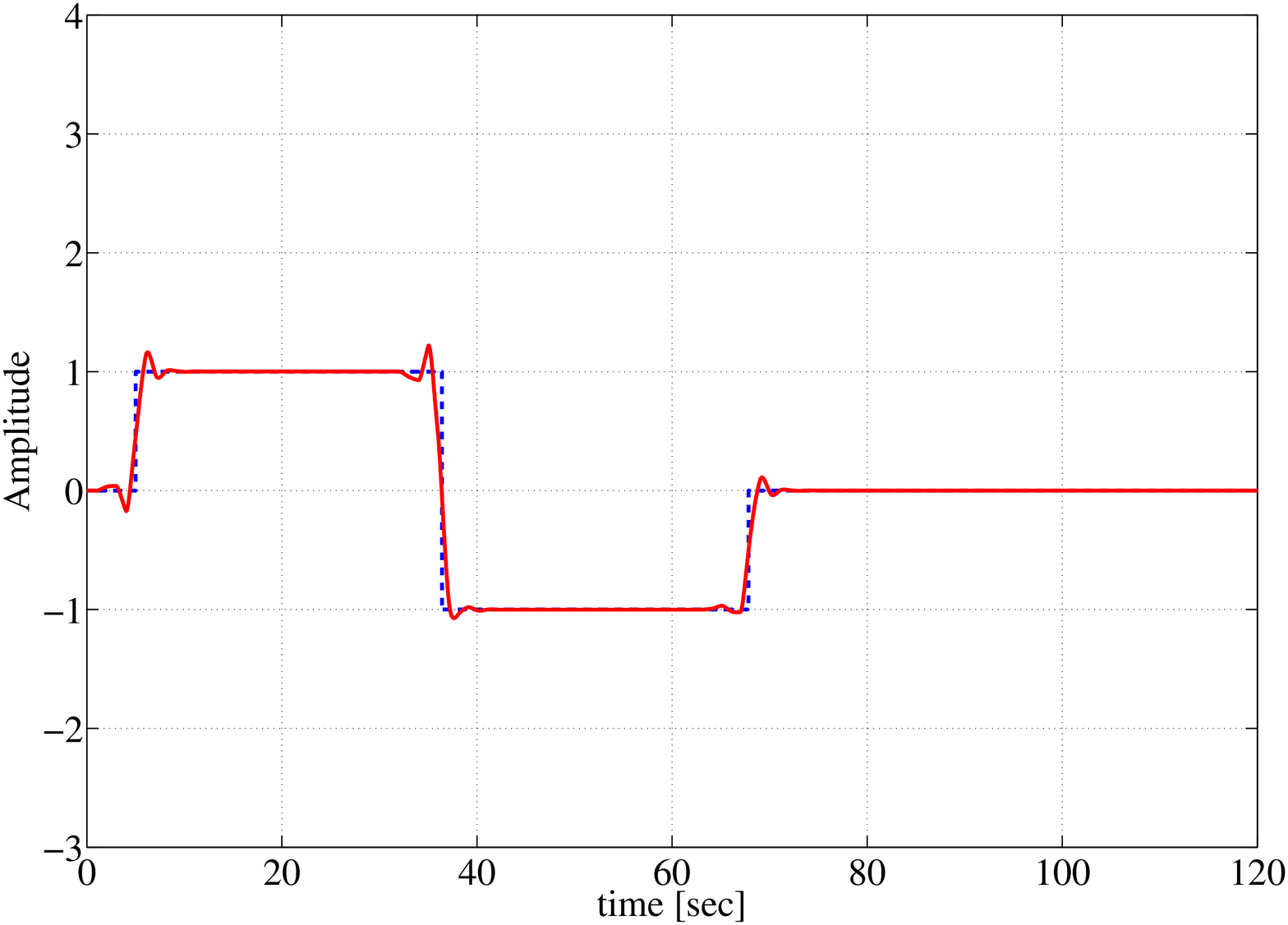}
   \caption{Reconstruction of a rectangular wave by the proposed method.}
   \label{fig:yy_reconstruction}
  \end{minipage}
 \end{tabular}
\end{figure}

Let us see this by an example. Consider 
\[
  \Fa(s)=\frac{1}{s+1}, \quad P(s) = \frac{1}{(s+d)(s+2)},\quad d=1.5.
\]
The zeros of the discretized system $H_{d}(z)$ are
\[
  \{-1.28549, -0.0816767\},
\]
and hence the optimal filter
\[
K_{\text{op}}(z) = \frac{1}{H_{d}(z)}
 = \frac{z^3 - 0.7263 z^2 + 0.1621 z - 0.01111}{0.05725 z^2 + 0.07827 z + 0.006011}
\]
has an unstable pole at $z=-1.28549$.
This $K_{\text{op}}$ agrees exactly with 
the one obtained via oblique projections; 
see, e.g., \cite{UnserWeb}.  
To implement this filter, 
we first shift $K_{\text{op}}(z)$ to obtain a proper transfer function,
that is, we use $K_{\text{op}}(z)=\frac{1}{zH_d(z)}$,
as suggested in \cite[Subsection V-C]{Unser05}.  
Then, we split it into two parts: 
the causal and anti-causal part.  
Fig.~\ref{fig:impulse_c} and Fig.~\ref{fig:impulse_ac} 
show the impulse responses of the causal and anti-causal part.
The anti-causal impulse response exhibits much oscillation 
because the pole $-1.28549$ is close to the point $z=-1$.  
Using this filter, we reconstruct a response against 
a rectangular wave.  
Fig.~\ref{fig:unser_reconstruction} shows the result.  
The reconstructed signal exhibits much oscillation around the edges.  
This is due to the oscillatory impulse response 
shown in Fig.~\ref{fig:impulse_ac}.
Moreover, the result shows a rather long delay, about 31 steps.  

Let us design the (sub)optimal filter $K_{\text{sd}}(z)$ 
by our sampled-data $H^\infty$ optimization method.  
Assume the analog characteristic of the input signals to be 
\[
 F(s) = \frac{1}{s+0.05}.
\]
We also assume the reconstruction delay $m=5$.  
Fig.~\ref{fig:yy_reconstruction} shows the reconstructed signal 
against the same rectangular wave.  
It is clearly seen that the proposed method 
provides a much better result.  

\begin{figure}[tbp]
 \centering
  \includegraphics[width=0.48\linewidth]{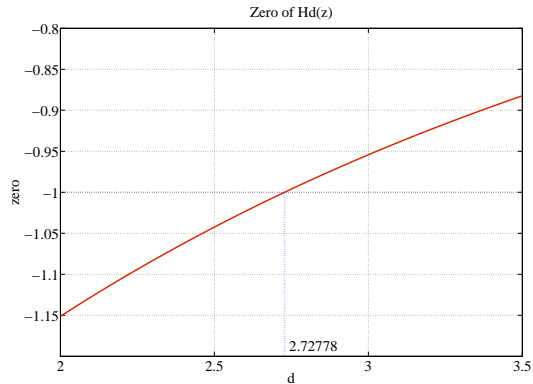}
 \caption{A zero in $K_{\text{op}}(z)$ versus the parameter $d$ in the post filter $P(s)=1/(s+2)(s+d)$.}
 \label{fig:unstable_zero}
\end{figure}

Let us further discuss the stability issue.  
It is claimed in \cite{Unser05} that there will be 
no zeros on the unit circle in the optimal filter, 
but this does not hold.  
Fig.~\ref{fig:unstable_zero} shows the locus of a pole of 
$K_{\text{op}}(z)$ for $d\in [2,3.5]$.  
There exists a real number $d$ (approximately $2.72778$) such that 
the filter $K_{\text{op}}(z)$ has a pole at $z=-1$.  
This clearly shows that the filter cannot be implemented 
as it is, and the optimal filter is not stable.  

\begin{figure}[tbp]
 \centering
  \includegraphics[width=0.48\linewidth]{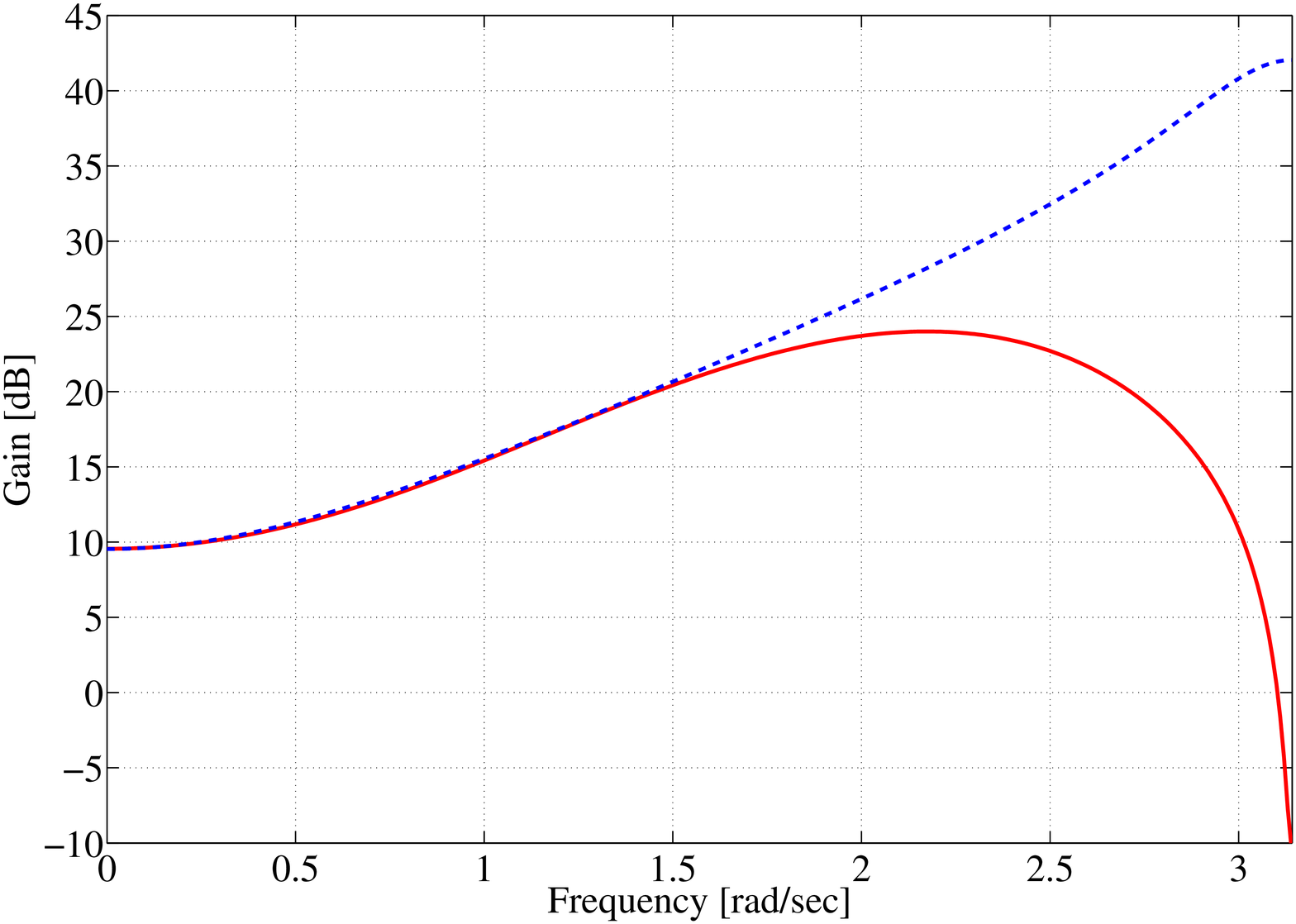}
 \caption{Frequency response of the filters $K_{\text{op}}(z)$ (dash) and $K_{\text{sd}}(z)$ (solid).}
 \label{fig:filter_gain}
\end{figure}

Fig.~\ref{fig:filter_gain} shows the frequency response of the filters $K_{\text{op}}(z)$ and $K_{\text{sd}}(z)$.

The optimal filter $K_{\text{op}}(z)$ shows a higher gain 
than the sampled-data filter $K_{\text{sd}}(z)$ in high frequency, 
and this explains the ringing around the edges 
in Fig.~\ref{fig:unser_reconstruction}.

\subsection{Comparison with the results of Meinsma and Mirkin}
\label{Subsec:MM}

Recently, Meinsma and Mirkin \cite{MeinsmaMirkinSP10a,MeinsmaMirkinSP10b} 
have also studied the signal reconstruction problem in a framework that is similar to ours. 
Some key features of their results are the following: 
\begin{itemize}
\item Solutions are given to 
signal reconstruction with 
 free sampler or free hold, or
 both free sampler and hold.

\item Instead of  a fixed preview length, 
they allow non-causal filters. 
\item An $L^{\infty}$ bound for the error performance as well as
$L^{2}$ (or $H^{2}$) type closed-form solutions are given.
\end{itemize}

Our work which started in 1995 has the following features:
\begin{itemize}
\item fixed sampler and hold devices, 
\item fixed preview length, and 
\item causal and stable $H^{\infty}$-optimal 
(suboptimal) filter construction.
\end{itemize}

Our motivation is that the fixed sampler and hold is 
 a more commonly encountered and practical situation.  
We have also chosen a fixed preview length ($e^{-mhs}$ in 
Fig.~\ref{Fig:err_interp}) as a design parameter, 
rather than considering optimal filter design 
with non-causal filters.  In the latter, when we 
have to implement it in practice, we need to 
truncate the impulse response at a certain length.  
A priori estimate of the
resulting performance after truncation is difficult to obtain.  When 
one imposes certain filter characteristics 
(e.g., fast decay beyond the pass-band), the amount of 
delays necessary to implement such filters can be 
very large, often reaching thousands of steps, 
as is the case of FIR filters having a sharp cut-off
characteristic.  On the other hand, once we fix 
a delay length, which corresponds to the preview length, 
our design Problem \ref{prob:Hinf} in 
Section \ref{Sec:SignalReconstruction} always gives 
rise to an optimal performance bound $\gamma$ 
(cf.\ (\ref{eqn:perfindex1})).  
Moreover, due to the very nature of sampled-data 
$H^{\infty}$ control, this will always yield a 
{\em stable\/} filter.   

\section{Applications to sound and image processing}
\label{Sec:Appl}

We here present some applications of the proposed method 
to sound and image processing.

\subsection{Application to sounds}

As seen in Example \ref{Ex:Johnston-yy} in Section \ref{Sec:FSFH},
the sharp cut-off characteristic based on the Shannon paradigm
generally induces a high distortion due to the Gibbs phenomenon.  

\begin{figure}[tbp]
 \centering
  \includegraphics[width=0.48\linewidth]{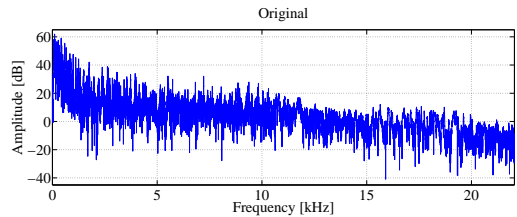}
 \caption{Fourier transform of original sound}
 \label{fig:orig_sound}
\end{figure}

\begin{figure}[tbp]
 \begin{tabular} {cc}
  \begin{minipage} {0.48\hsize}
    \includegraphics[width=\linewidth]{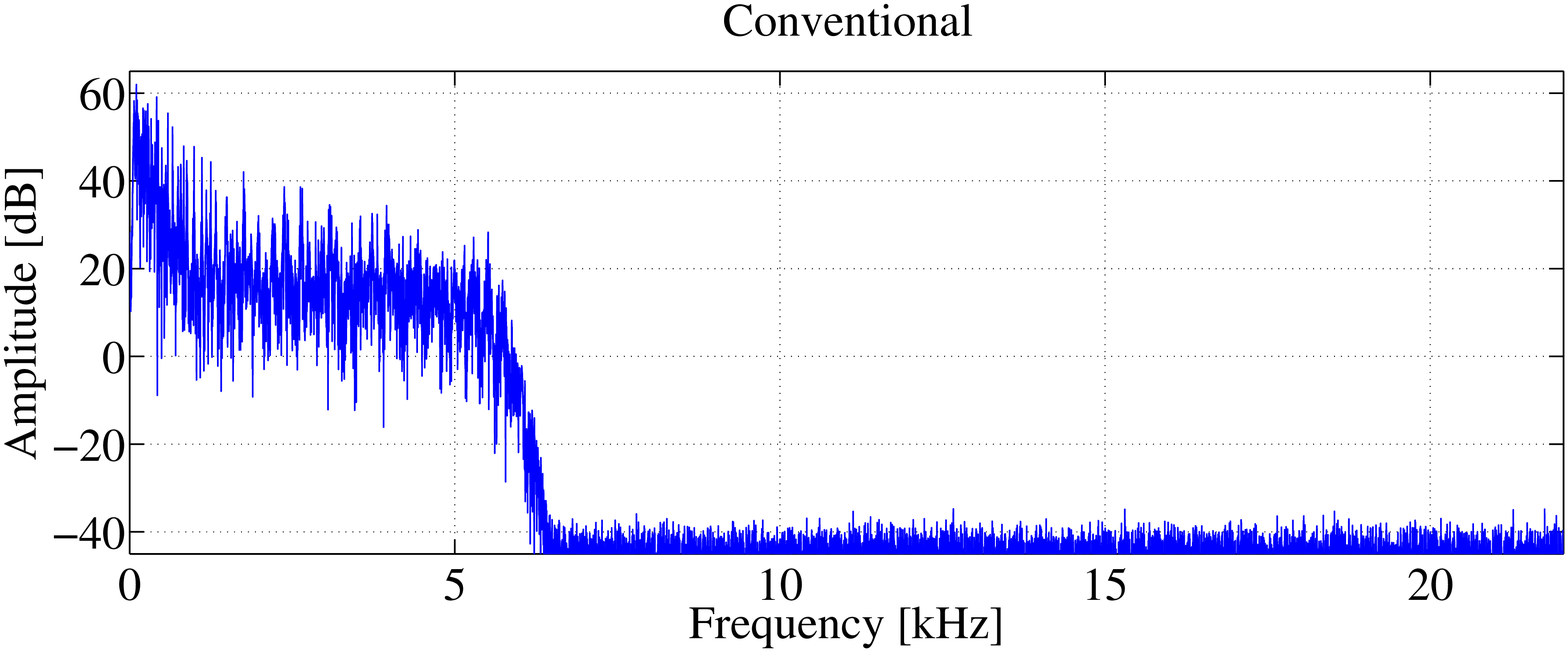}
    \caption{FFT of reconstructed sound by equi-ripple filter~~~~~~~~}
    \label{fig:equi_sound}
  \end{minipage}
  \begin{minipage} {0.48\hsize}
   \includegraphics[width=\linewidth]{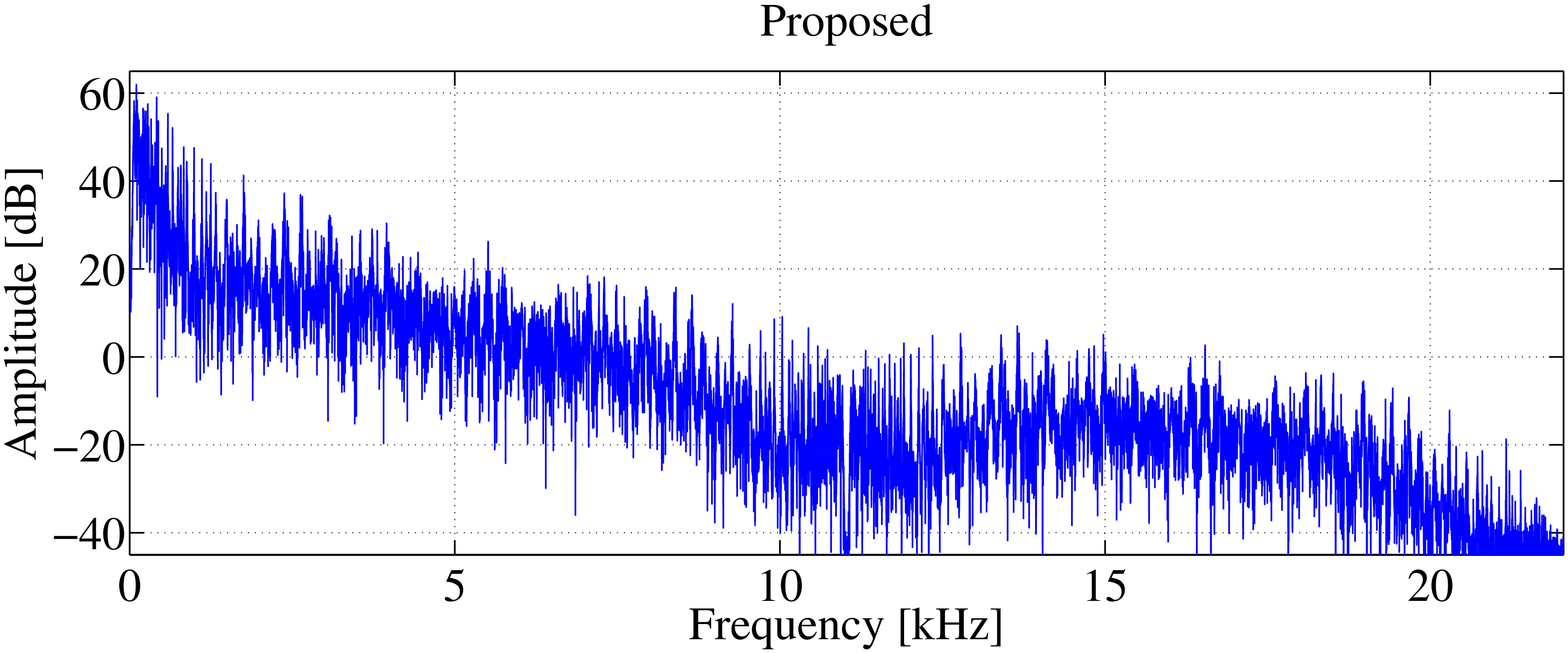}
   \caption{FFT of reconstructed sound by sampled-data filter}
   \label{fig:yy_sound}
  \end{minipage}
 \end{tabular}
\end{figure}

We here show a brief example of sound recovery in high frequency.
We consider a sound sample whose Fourier transform is shown in 
Fig.~\ref{fig:orig_sound}.  
This signal has the frequency components up to $22.05$ kHz.  
We apply downsampler $\ds{4}$ to this signal 
to obtain a signal whose bandwidth is limited to 5.51 kHz, and 
attempt to recover the original high frequency components.  
We upsample it by the factor of 4, and then 
apply two filters: a conventional equi-ripple filter and the proposed filter.
Fig.~\ref{fig:equi_sound} shows the Fourier transform of 
the reconstructed signal by the equi-ripple filter, while 
Fig.~\ref{fig:yy_sound} shows the Fourier transform
of the recovered sounds by the proposed method, with a 
suitable $F(s)$ as in Example \ref{Ex:Johnston-yy}.  
We can see that high frequency components beyond 6 kHz are 
well recovered by the sampled-data method.
On the other hand, if we apply the 
equi-ripple filter with cut-off frequency 5.51 kHz, 
it does not give any frequency components beyond 6 kHz, 
naturally, since there is no guiding principle for 
reconstructing such components beyond 5.51 kHz.  
The advantage of the present method is that it can 
evaluate the overall performance of the error signal 
$e_{c}$ in Fig.~\ref{Fig:err_interp} in terms of 
the $\Hinf$-norm of $T_{ew}$ of the transfer operator.  


The present method has been applied to sound processing, 
particularly in supplementing lost high-frequency components 
in compression audio, and has been quite successful.  
In sound compression, the bandwidth is often limited 
to a rather narrow range (e.g., only up to 12 or 16 kHz,
as in the MP3 or AAC format).  
The digital filter using sampled-data theory here, 
along with upsampling, can recover the lost intersample 
information optimally in the $H^{\infty}$ sense, thereby 
expanding the effective bandwidth to the original range.  
This has been patented \cite{YYPatent1,YYPatent2,SANYOPatent1,SANYOPatent2} 
and implemented into sound processing LSI 
chips as a core technology by Sanyo Semiconductors, and successfully 
used in mobile phones, digital voice recorders and MP3 players; 
their cumulative production has exceeded 
25 million units as of 2011. 

\subsection{Application to Images}
\label{Sec:Appl:Images}
The same idea is applicable to images.  
However, since images are two-dimensional, we should 
be careful about how our (essentially) one-dimensional 
method can be applied.  There is no universal recipe for this,
and the simplest is to apply this in two steps: first process 
the data in the 
horizontal direction, save the temporary data in buffer memories,
and then process them in the vertical direction.  

\begin{figure}[tbp]
 \begin{tabular} {cc}
  \begin{minipage} {0.48\hsize}
   \centering
    \includegraphics[width=0.8\linewidth]{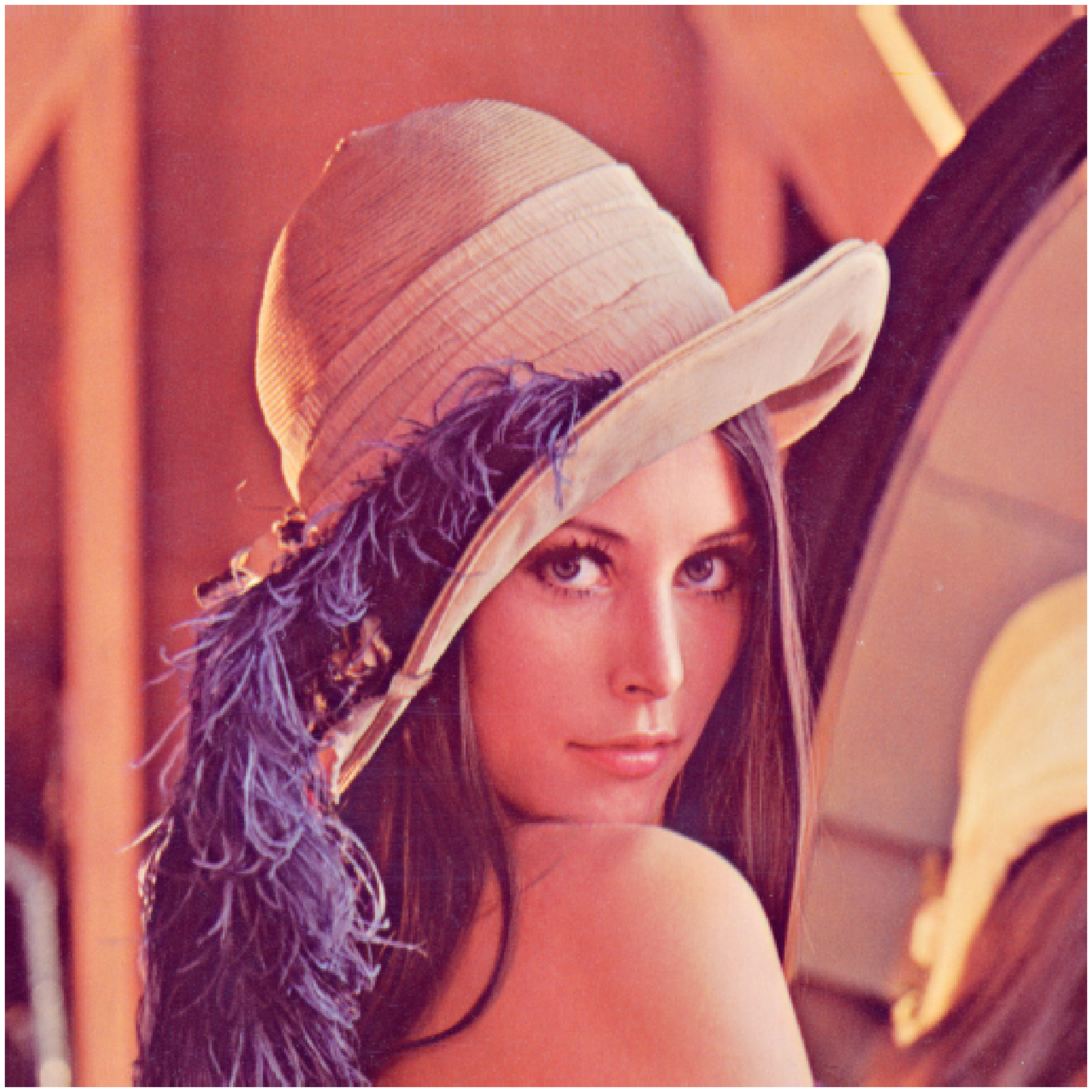}
    \caption{Lena}
    \label{fig:lena}
  \end{minipage}
 \begin{minipage} {0.48\hsize}
  \centering
    \includegraphics[width=0.8\linewidth]{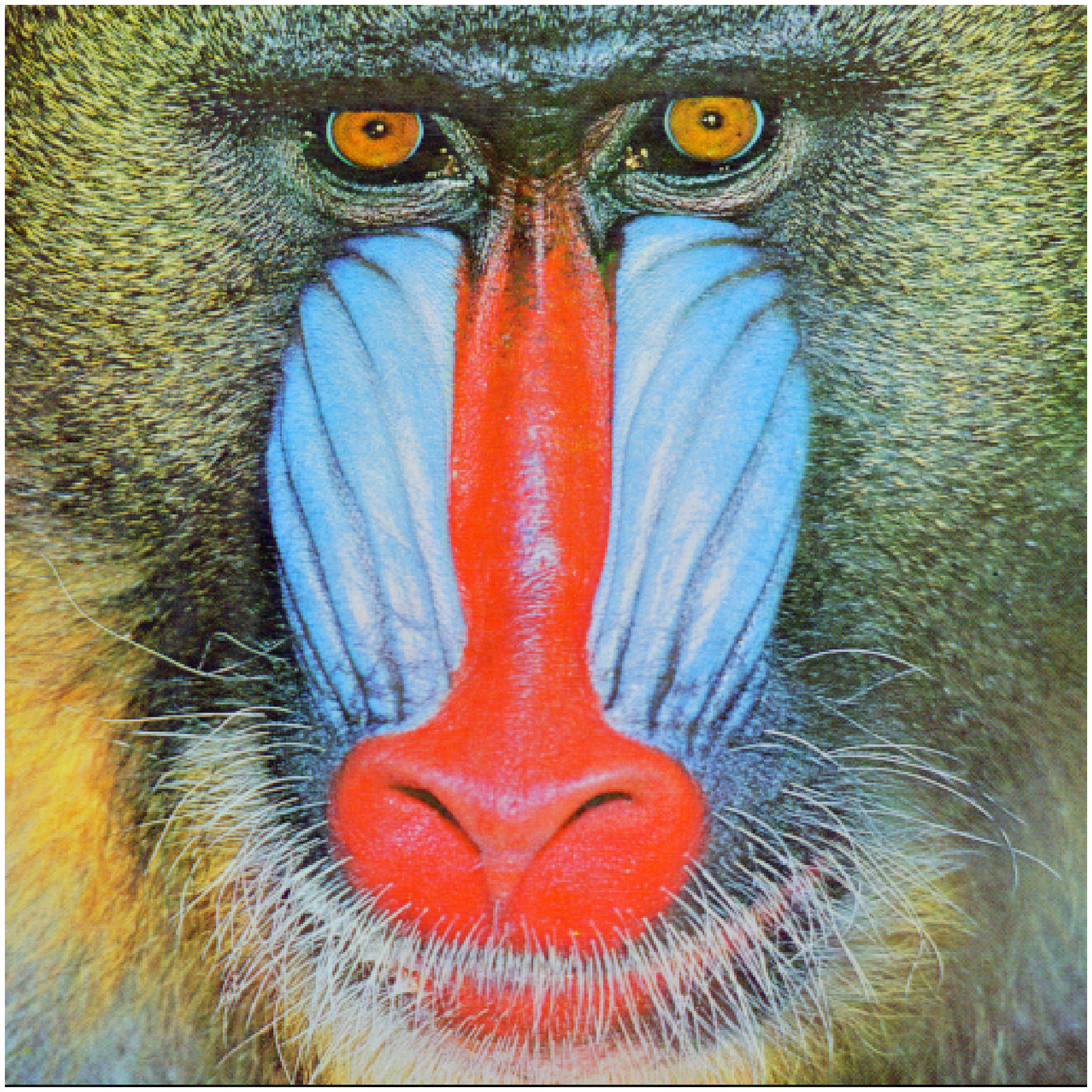}
    \caption{Baboon}
    \label{fig:baboon}
 \end{minipage}
\end{tabular}
\end{figure}

\begin{figure}[tbp]
 \centering
  \mbox{
   \subfigure[Downsampled image]{
    \includegraphics[width=0.3\linewidth]{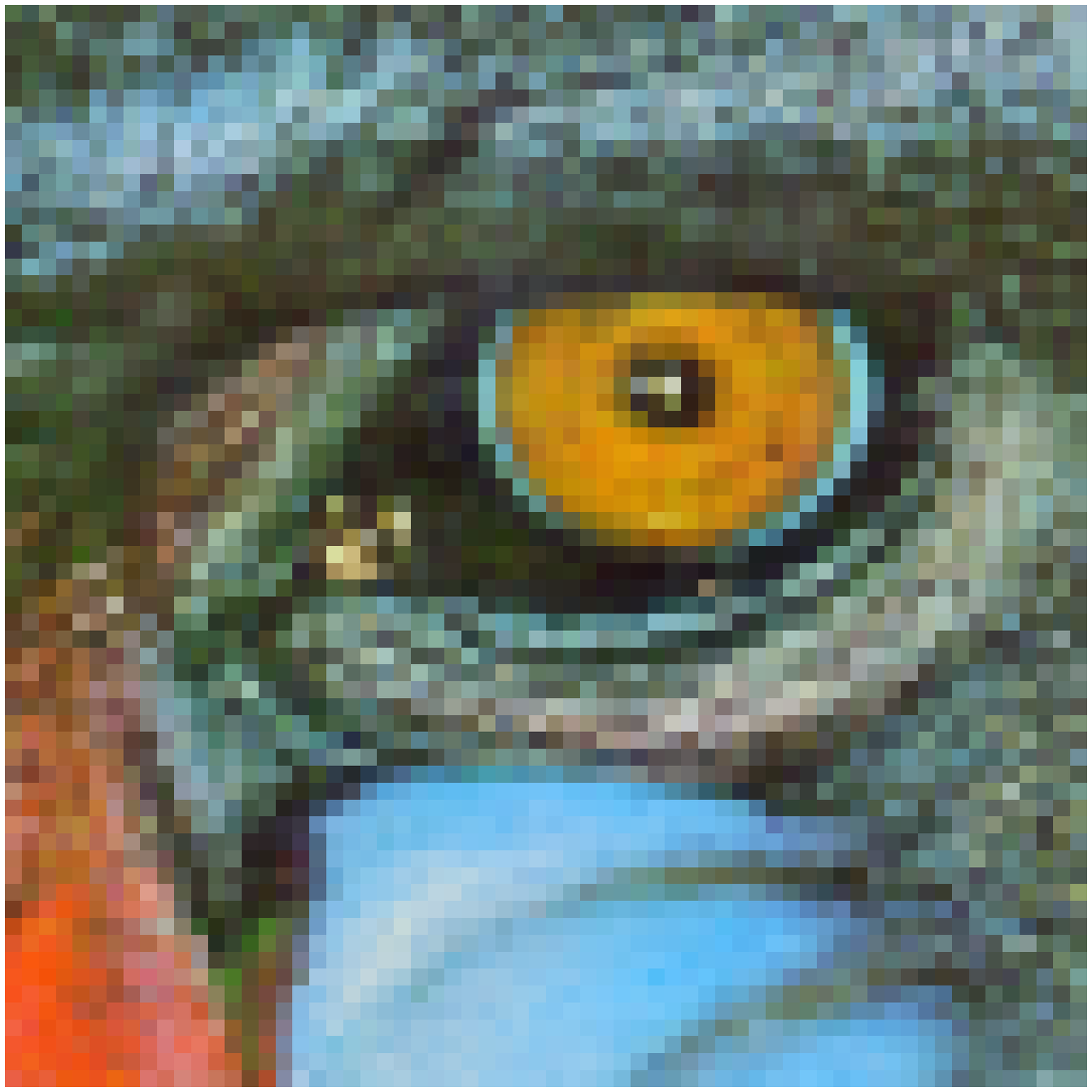}
    \label{fig:baboon_d}
   }\quad
   \subfigure[Lanczos method]{
    \includegraphics[width=0.3\linewidth]{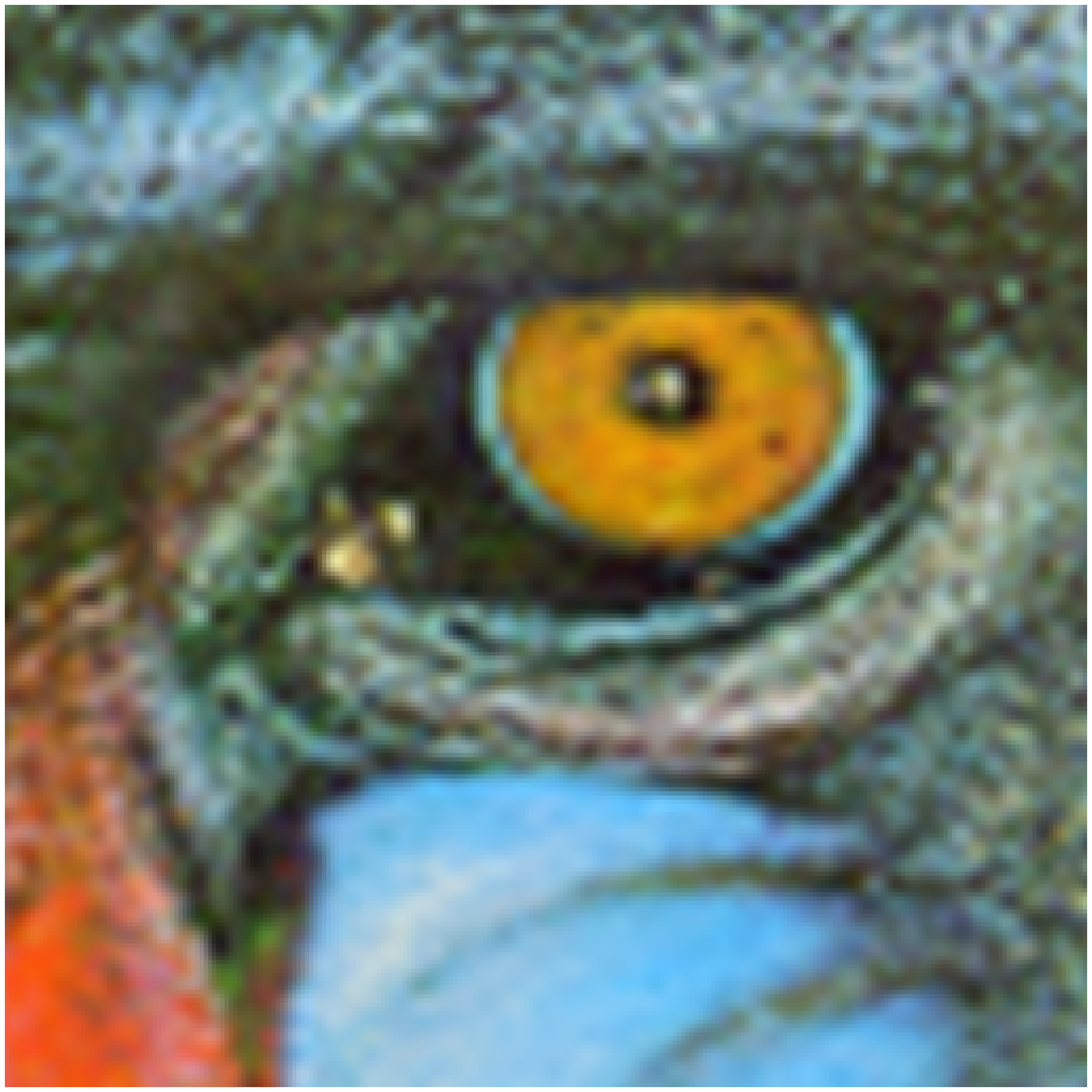}
    \label{fig:baboon_l}
   }}\\
   \mbox{
   \subfigure[TV method]{
    \includegraphics[width=0.3\linewidth]{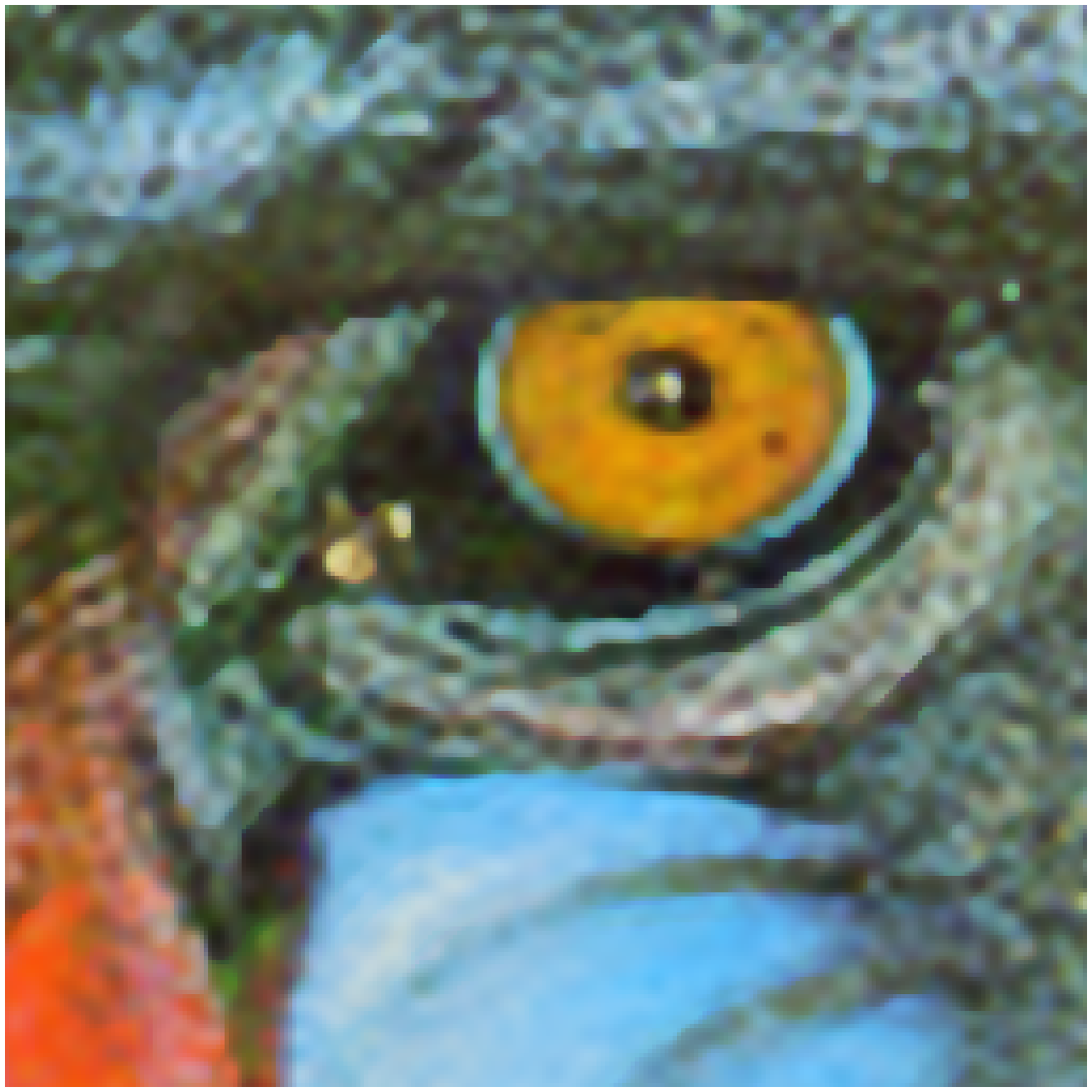}
    \label{fig:baboon_tv}
   }\quad
   \subfigure[Proposed method]{
    \includegraphics[width=0.3\linewidth]{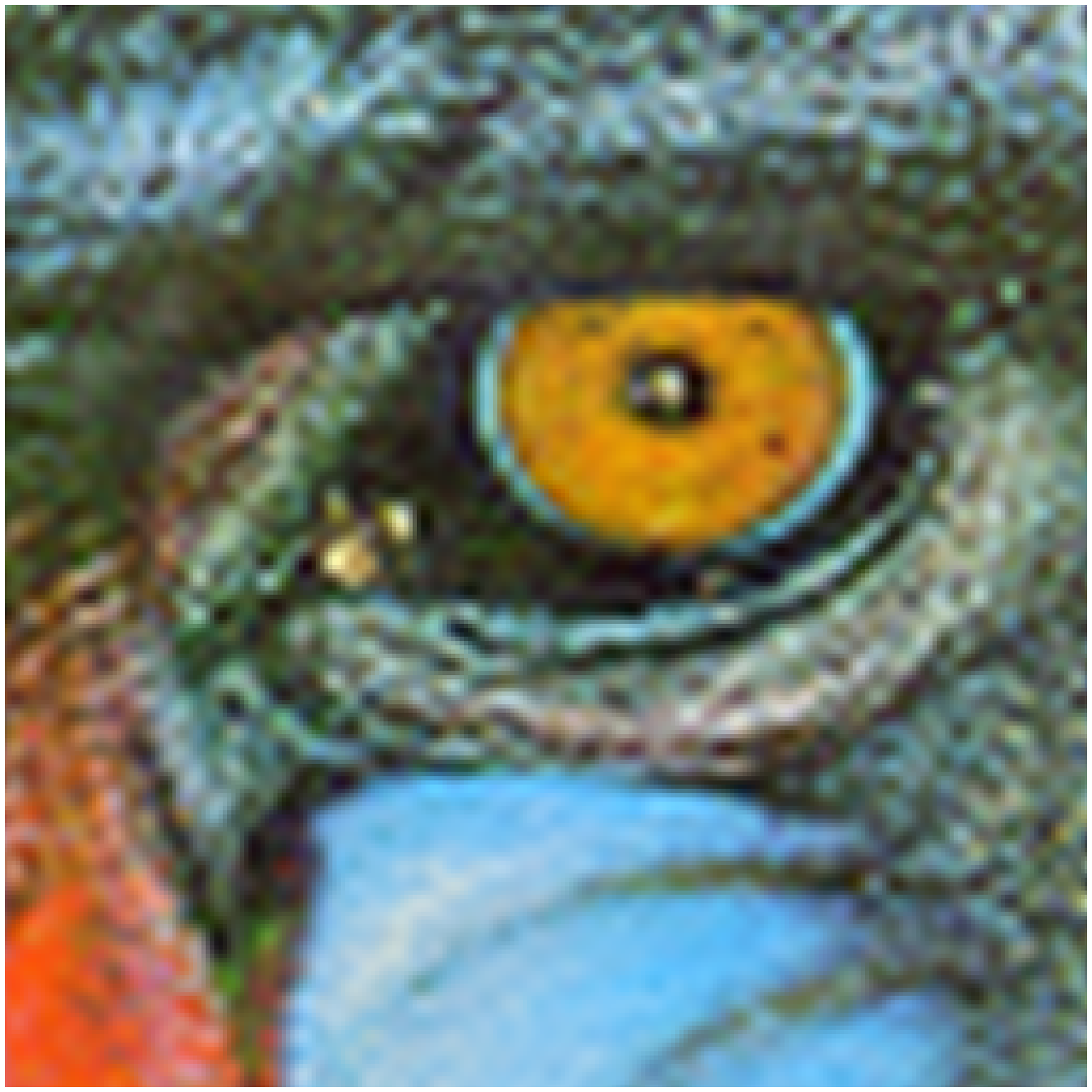}
    \label{fig:baboon_yy}
   }
  }
  \caption{Image processing results}
  \label{fig:baboons}
\end{figure}

We can interpolate lost intersample data 
by the present framework.  For example, 
take the well-known sample picture of 
Lena shown in Fig.~\ref{fig:lena},
and Baboon shown in Fig.~\ref{fig:baboon}.
We downsample it to an image of size $1/4$.
Then from the downsampled image we attempt to reconstruct
the original image via
the Lanczos method \cite{Lanczos},
total variation (TV) regularization method \cite{TV},
and the proposed sampled-data $H^\infty$ method.
The Lanczos method uses a windowed sinc filter and is based on the sampling theorem.
The TV criterion penalizes the total amount of change in the image
to preserve steep local gradients.
This method is very popular in super-resolution imaging.
Table \ref{tbl:2x_psnr} shows the reconstruction performances
measured by their peak signal-to-noise ratio (PSNR).
\begin{table}[tbp]
\caption{PSNR performance (dB)}
\label{tbl:2x_psnr}
\begin{center}\begin{tabular}{cccc}
	\hline
	Image & Lanczos & TV & {\bf Proposed}\\
	\hline
	Lena   & 33.5497 & 33.3760 & {\bf 33.6748} \\
	Baboon & 23.2691 & 23.2303 & {\bf 23.2813} \\\hline
\end{tabular}
\end{center}
\end{table}
For these two images, the proposed method shows the best performance.
We also show the performance measured by their structural similarity (SSIM)
\cite{SSIM}.
\begin{table}[tbp]
\caption{SSIM performance}
\label{tbl:2x_ssim}
\begin{center}
\begin{tabular}{cccc}
	\hline
	image & Lanczos & TV & {\bf Proposed} \\
	\hline
	Lena   & 0.9962 & 0.9962 & {\bf 0.9992} \\
	Baboon & 0.9881 & 0.9893 & {\bf 0.9989} \\\hline
\end{tabular}
\end{center}
\end{table}
Also in SSIM, the proposed method shows the best performance.
Figs.~\ref{fig:baboon_d} -- \ref{fig:baboon_yy} show the results.
Fig.~\ref{fig:baboon_d} is the downsampled image.
From this image, we reconstruct the original image.
Fig.~\ref{fig:baboon_l} is the reconstructed image by the Lanczos method,
Fig.~\ref{fig:baboon_tv} by the TV method, and
Fig.~\ref{fig:baboon_yy}, by the proposed method.
The Lanczos method produces a blurred image since this method
is based on the sampling theorem.
As a result,
high frequency components
are discarded by the windowed sinc filter.
The reconstructed image by the TV method has artificial edges,
in particular in the edge between the eyelid and the pupil.
Since the TV method attempts to reduce delicate changes and
preserve steep gradients, the reconstructed image
appears as a painting.
Compared with these results, the proposed method shows
an accurate reconstruction; the details of the skin
around the eye are well recovered.
Note that the TV method uses an iteration in computing the image,
which makes it more demanding computationally
than the proposed method
which is just linear filtering.

\section{Concluding Remarks} 

We have presented a new framework for digital 
signal processing.  The fundamental philosophy is 
the emphasis on analog (continuous-time) performance with
 discrete-time signal processing.  This naturally 
leads to a technical difficulty because two different 
time-sets are involved: continuous and discrete.  
Leveraging the sampled-data $H^{\infty}$ control theory,
we have presented computable procedures for designing 
optimal, stable, causal filters. These filters are optimal with respect to a uniform analog performance measure. 
Our methodology 
is applicable to a wide variety of theoretical and 
application problems in digital signal processing.  We believe that it has many merits and we hope it will be more widely used in the future. 

\section*{Acknowledgments}
The authors would like to thank Professor Bruce Francis for his valuable
comments, and also Mr.\ Kengo Zenitani for the computation of image
examples.  This research is supported in part by the JSPS Grant-in-Aid
for Scientific Research (B) No. 2136020318360203, and also
by Grant-in-Aid for Exploratory Research No. 22656095.  Pramod
Khargonekar is also supported in part by Eckis Professor endowment.

\appendices

\section{Lifting, Transfer Functions, and Frequency Responses}
\label{Sec:appendLifting}

As mentioned in the main text, the major difficulty in sampled-data
systems lies in the mixture of two different time
sets: continuous and discrete.  
{\em Lifting\/} \cite{BamiehPearson92,ChenFrancisBook,YYTAC94,YYencyclopedia} 
is a method that makes it possible to describe continuous-time 
systems in a discrete-time setting, without introducing any 
approximation, thereby merging the two time sets into one.  

\begin{figure}[tb]
\centering
\includegraphics[width=0.48\linewidth]{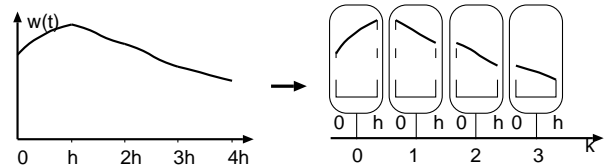}
\caption{Lifting: a continuous-time signal $w(t)$ (left) is converted to a function-valued discrete time signal (right)}
\label{fig:lifting}
\end{figure}

We start by placing a continuous-time signal in a discrete-time 
framework.  Take a continuous-time signal $w(t)$, and consider 
the following mapping $\cal L$ (with a suitable domain and codomain)
that maps $w$ into a {\em sequence of functions\/} as 
\begin{equation} \label{eq:lifting0}
(\mathcal{L} w)[k] := \boldsymbol{w}[k] := \{w(kh + \theta)\}_{\theta\in[0,h)},\ k=0,1,2,\ldots.
\end{equation}
See Fig.~\ref{fig:lifting}.
The operator ${\mathcal L}$ is called {\em lifting}.  

This idea makes it possible to view time-invariant, or even periodically time-varying 
continuous-time systems as linear, time-invariant discrete-time systems.

Using this operator, one can describe a linear, time-invariant continuous-time
system with a linear, time-invariant discrete-time system.  
Consider the following linear system: 
\begin{equation} \label{eqn:LTIsysappdx}
 \begin{split}
  \dot{x}(t) &= A x(t) + B u(t),\\
  y(t) &= C x(t),
 \end{split}
\end{equation}
where $x(t) \in \R^{n}$, $u(t) \in \R^{m}$ and 
$y(t) \in \R^{p}$ are the state, input and output 
of this system, respectively.  
Let us assume, e.g., $u\in L^2_{\rm{loc}}[0,\infty)$, 
the set of locally square-integrable functions on $[0,\infty)$.  
The idea is that we view the continuous-time system (\ref{eqn:LTIsysappdx}) 
as one with discrete-timing $t = kh$, $k=0,1,2,\ldots$ 
such that it receives function-valued inputs at these instants and
produces function-valued outputs at these times also.  
Suppose that (\ref{eqn:LTIsysappdx}) is at state $x(kh)$ at time 
$t = kh$.  Then 
\[
 \begin{split}
  x((k+1)h) &= e^{Ah}x(kh) + \int_{0}^{h} e^{A(h-\tau)}Bu(kh+\tau)d\tau \\
  y(kh+\theta) &= C e^{A\theta}x(kh) + 
          \int_{0}^{\theta} e^{A(\theta-\tau)}Bu(kh+\tau)d\tau.
 \end{split}
\]
where $0 \leq \theta < h$ denotes the intersample parameter.  
Lifting the input $u(t)$ and the output $y(t)$ as per (\ref{eq:lifting0}), 
we can rewrite these formulas as 
a lifted discrete-time system \cite{ChenFrancisBook, YYencyclopedia}:
\[
 \begin{split}
  x[k+1] &= {\mathcal A} x[k] + {\mathcal B} \boldsymbol{u}[k],\\
  \boldsymbol{y}[k] &= {\mathcal C} x[k] + {\mathcal D} \boldsymbol{u}[k],\quad k=0,1,2,\ldots
 \end{split}
\]
where $x[k]=x(kh)$, $\boldsymbol{u}[k]=({\mathcal L}u)[k]$, $\boldsymbol{y}[k]=({\mathcal L}y)[k]$, and
\begin{equation} \label{eqn:Liftedsystem}
 \begin{aligned}
 {\mathcal A}&: {\mathbb R}^{n} \rightarrow {\mathbb R}^{n}: x \mapsto e^{Ah}x,\\
 {\mathcal C}&: {\mathbb R}^{n} \rightarrow L^2[0,h): x \mapsto Ce^{A\theta}x,
 \end{aligned}
 \begin{aligned}\quad
 {\mathcal B}&: L^2[0,h) \rightarrow {\mathbb R}^{n}: u \mapsto \int_0^he^{A(h-\tau)}Bu(\tau)d\tau\\
 {\mathcal D}&: L^2[0,h) \rightarrow L^2[0,h): u \mapsto \int_0^\theta Ce^{A(\theta-\tau)}Bu(\tau)d\tau,
 \end{aligned}
\end{equation}
where $\theta\in[0,h)$ describes the intersample parameter. 
Observe that 
the operators $\cal A, B, C, D$ above do not depend on 
time $k$, and hence system (\ref{eqn:Liftedsystem}) is a 
{\em time-invariant discrete-time\/} 
system, albeit with infinite-dimensional input
and output spaces. 
Hence it is straightforward to connect this system with a 
discrete-time controller (or a filter), and the obtained 
sampled-data system is again a linear, time-invariant 
discrete-time system {\em without sacrificing any intersampling information}.  
The resulting system can also be described by a 4-tuple of operators
$\cal A, B, C, D$, and its 
{\em transfer function (operator)\/} of the lifted system is defined as 
\[
 G(z) = {\mathcal D} + {\mathcal C}(zI-{\mathcal A})^{-1}{\mathcal B}.
\]
with such $\cal A, B, C, D$.  
Note that for each fixed $z\in{\mathbb C} \setminus \sigma(e^{Ah})$, 
($\sigma(e^{Ah})$:= the set of eigenvalues of $e^{Ah}$), 
$G(z)$ is a linear operator acting on $L^2[0,h)$ into itself.  
The {\em frequency response operator\/} 
is then defined as $G(e^{j\omega h})$, 
and the {\em gain\/} at frequency $\omega$ is defined as 
\[
 \|G(e^{j\omega h})\| = \sup_{\begin{subarray}{c}v\in L^2[0,h)\\v\neq 0\end{subarray}}\frac{\left\|G\left(e^{j\omega h}\right)v\right\|}{\|v\|}.
\]
The $H^\infty$ norm of $G$ then becomes 
\[
 \|G\|_\infty = \sup_{\omega\in[0,2\pi/h)} \|G(e^{j\omega h})\|.
\]
which is known to be identical to the $L^2$-induced norm given by 
(\ref{eqn:SDinducednorm}) in Section \ref{Sec:preliminaries} 
\cite{ChenFrancisBook}.  

\section{Proof of Theorem \ref{th:fsfh}}
\label{Sec:appendFSFH}

Let ${\mathcal T}_N$ be the fast discretization shown in 
Fig.~\ref{fig:fsfh_fig}, namely,  
\[
  {\mathcal T}_N = \samp{h/N}T_{ew}\hold{h/N}
   = \samp{h/N}e^{-mhs}F\hold{h/N} - \samp{h/N}P\ghold{h}\widetilde{K}\samp{h}F\hold{h/N}.
\]
By using the identities
\[
  \ghold{h}=\hold{h/N}\idlift{N}H,\quad \hold{h/N}\samp{h/N}\hold{h/N} = \hold{h/N},\quad
  \samp{h}=S\dlift{N}\samp{h/N},
\]
where
\[
  H := \diag{\{I_l\}}\in\R^{N\times L},\quad I_l:=[1,1,\ldots ,1]^T\in\R^{l},\quad
  S := [1,0,\ldots,0] \in \R^{1\times N},
\]
we have
\[
 {\mathcal T}_N = z^{-mN}{\mathcal F}_N - {\mathcal P}_N\idlift{N} H \widetilde{K} S\dlift{N} {\mathcal F}_N,
\]
where ${\mathcal F}_N := \samp{h/N}F\hold{h/N}$ and ${\mathcal P}_N := \samp{h/N}P\hold{h/N}$.
Applying the discrete-time lifting $\dlift{N}$ and its inverse $\idlift{N}$ gives
\[
  T_N = \dlift{N}{\mathcal T}_N\idlift{N}
   = z^{-m}\dlift{N}{\mathcal F}_N\idlift{N} - \dlift{N}{\mathcal P}_N\idlift{N} H \widetilde{K} S\dlift{N} {\mathcal F}_N\idlift{N}
   = z^{-m}F_N - P_N H \widetilde{K} S F_N.
\]
The state space matrices for $P_N$ and $F_N$ are given by the formulas 
in Theorem 8.2.1 \cite[Chap.~8]{ChenFrancisBook}.
The convergence in (\ref{eq:fsfh-convergence}) 
is proved in \cite{YYAutoma99} and in \cite{YYCDC02}.  It is 
uniform in frequency \cite{YYAutoma99}, and also 
uniform in $\widetilde{K}$ when the filter is confined to 
a compact set \cite{YYCDC02}.

\section{Proof of Theorem \ref{th:inverse}}
\label{Sec:appendProofThm1}

First, we consider the denominator of $K_{\text{op}}(z)$.
The coefficients $\beta(k)$, $k=0,1,2,\dots$ are
obtained by sampling the inverse Laplace transform of $\hat{\beta}$ given in (\ref{eq:laplace_beta}).
Since $(1-e^{-s})/s$ is the Laplace transform of the zero-order hold 
$\mathcal{H}$, 
the denominator of $K_{\text{op}}(z)$ is the $Z$-transform of
the step-invariant transformation of
\[
 \tilde{\beta}(s):=\prod_{i=1}^{2}\prod_{n=1}^{n_i}(1-e^{\alpha_{in}}e^{-s})\Fa(s)P(s).
\]
That is, the denominator is given by
\[
  \mathcal{Z}\!\left[\samp{}\tilde{\beta}(s)\hold{}\right]
   =\mathcal{Z}\!\left[\samp{}\!\left(\prod_{i=1}^{2}\prod_{n=1}^{n_i}(1-e^{\alpha_{in}}e^{-s})\Fa(s)P(s)\right)\!\hold{}\right].
\]
By using the relation $\samp{}e^{-s}=z^{-1}\samp{}$ and the definition of $\Delta_i(z)$
in (\ref{eq:dlocalization}),
we have
\[
\samp{}~\prod_{i=1}^2\prod_{n=1}^{n_i}(1-e^{\alpha_{in}}e^{-s})
=\Delta_{1}(z)\Delta_{2}(z)\samp{}.
\]
It follows that
\[
\samp{}\tilde{\beta}(s)\hold{}
=\Delta_{1}(z)\Delta_{2}(z)\samp{}\Fa(s)P(s)\hold{}
=\Delta_{1}(z)\Delta_{2}(z)H_{d}(z).
\]
Then, since the numerator of $K_{\text{op}}(z)$ is $\Delta_{1}(z)\Delta_{2}(z)$,
we conclude that
\[
K_{\text{op}}(z) = \frac{\Delta_{1}(z)\Delta_{2}(z)}
	{\mathcal{S} \tilde{\beta}(s)\mathcal{H}}
=\frac{1}{H_{d}(z)}.
\]

\end{document}